\documentclass[10pt,letterpaper]{article}
\usepackage[top=0.85in,left=1.6in,footskip=0.75in]{geometry}

\usepackage{amsmath,amssymb}

\usepackage{changepage}
\usepackage[shortlabels]{enumitem}

\usepackage[utf8x]{inputenc}

\usepackage{textcomp,marvosym}

\usepackage{cite}

\usepackage{nameref}
\usepackage{hyperref}

\usepackage[right]{lineno}

\usepackage{microtype}
\DisableLigatures[f]{encoding = *, family = * }

\usepackage[table]{xcolor}

\usepackage{array}

\usepackage{tabularx}
\usepackage{caption}
\usepackage{booktabs}

\usepackage{soul,color}   
\usepackage{float}            
\usepackage{cleveref}      
\usepackage{multirow}      
\usepackage {tabu}			   
\usepackage{longtable}    
\newcolumntype{C}[1]{>{\centering\let\newline\\\arraybackslash\hspace{0pt}}m{#1}}
\usepackage{graphicx}     
\usepackage{subcaption} 
\usepackage{wrapfig}       
\usepackage{soul}            
\usepackage{pdfpages}  
\usepackage{hyperref}  
\usepackage{xcolor}      

\usepackage{ifthen}
\usepackage{helvet}
\makeatletter
\newcommand{\showfont}{encoding: \f@encoding{},
  family: \f@family{},
  series: \f@series{},
  shape: \f@shape{},
  size: \f@size{}
}

\newcolumntype{+}{!{\vrule width 2pt}}

\newlength\savedwidth

\usepackage[aboveskip=1pt,labelfont=bf,font=small,labelsep=period,singlelinecheck=off]{caption}

\makeatletter
\renewcommand{\@biblabel}[1]{\quad#1.}
\makeatother

\usepackage{lastpage,fancyhdr,graphicx}
\usepackage{epstopdf}
\begin{document}
\vspace*{0.2in}

\begin{flushleft}
{\Large
\textbf\newline{\textbf{Step-wise target controllability identifies dysregulations of macrophage networks in multiple sclerosis}\\
}
}

\vspace{5mm}


Giulia Bassignana\textsuperscript{1,2},
Jennifer Fransson\textsuperscript{1},
Vincent Henry\textsuperscript{1,2},
Olivier Colliot\textsuperscript{1,2},
Violetta Zujovic\textsuperscript{1},
Fabrizio De Vico Fallani\textsuperscript{1,2,*}
\\
\bigskip
\textbf{1} Sorbonne Universites, UPMC Univ Paris 06, Inserm U-1127, CNRS UMR-7225, Institut du Cerveau et de la Moelle Epini\`{e}re, Hopital Piti\'{e}-Salp\^{e}tri\`{e}re, Paris, France\\
\textbf{2} Inria Paris, Aramis Project Team, Paris, France
\\

\bigskip

%
%
%



* corresponding author: fabrizio.devicofallani@gmail.com

\end{flushleft}
\bigskip

\section*{Abstract}
\label{sec: abstract}

\begin{small}

\bigskip

Identifying the nodes able to drive the state of a network is crucial to understand, and eventually control, biological systems. 
Despite recent advances, such identification remains difficult because of the huge number of equivalent controllable configurations, even in relatively simple networks.
Based on the evidence that in many applications it is essential to test the ability of individual nodes to control a specific target subset, we develop a fast and principled method to identify controllable driver-target configurations in sparse and directed networks.
We demonstrate our approach on simulated networks and experimental gene networks to 
to characterize macrophage dysregulation in human subjects with multiple sclerosis.

\end{small}
\newpage
\section{Introduction}
\label{sec: introduction}


For many biological systems, it is crucial to identify the units, such as genes or neurons, with the potential to influence the rest of the network, as this identification can enable describing, understanding, and eventually controlling the function of the system \cite{jeong_lethality_2001, bonifazi_gabaergic_2009}.
Topological descriptors based on network science can indeed be used to quantify such influence in terms of node centrality, such as degree, betweenness, or closeness \cite{newman_networks:_2010}.
However, these descriptors only capture the structural properties of the network and neglect their effect on the dynamics, thus limiting our understanding on the actual influencing power.

Control network theory, linking network structure to dynamics through linear or nonlinear models, has been shown to be a more principled approach for identifying driver nodes in an interconnected system  \cite{rugh_linear_1995, sontag_mathematical_1998}.
While theoretically these approaches can give a minimum set of driver nodes sufficient to steer the system into desired states, their exhaustive identification might be difficult in practice as there exists in general a very large number of equivalent controllable walks, even in relatively simple networks \cite{wagner_number_2007, heuberger_number_2011 }.
In the case of criteria based on the manipulation of controllability matrices \cite{kalman_mathematical_1963, hautus_stabilization_1970}, the presence of many walks can for example induce numerical errors due to the different orders of magnitude in the matrix elements. 
 
An alternative solution has recently been proposed to circumvent this limitation, based on the possibility to map the controllability problem onto the maximum cardinality matching over the associated graph \cite{lin_structural_1974, shields_structural_1976, liu_controllability_2011}.
As a result, it is possible to identify a set of driver nodes - at least for directed networks -  with linear, and not exponential, time complexity \cite{hopcroft_n5/2_1971}.
While this approach elegantly solves numerical issues, it can nevertheless not tell which configuration, among all the possible ones, is the most relevant. 
In general, there is a factorial number of equivalent configurations (with the same number of inputs) and enumerating all possible matchings \cite{uno_algorithms_1997} rapidly becomes unfeasible, even for simple graphs such as trees  \cite{wagner_number_2007, heuberger_number_2011}, bipartite graphs \cite{liu_number_2004}, or random graphs \cite{zdeborova_number_2006}. 
Thus, the research of alternative strategies to characterize the candidate driver nodes is crucial for the concrete application of network controllability tools.

One possibility would be to reduce the original problem into smaller sub-problems under the assumption of specific constraints compatible with the underlying scientific question. On the one hand, for many biological, technological and social systems it is desirable to only control a subset of target nodes (or a subsystem) that is essential for the system’s mission pertaining to a selected task or function. In this direction some approaches have been recently proposed, based on filtering of the controllability matrix \cite{klickstein_control_2019} or adaptation of graph matching \cite{gao_target_2014, czeizler_target_2016, zhang_efficient_2017, li_structural_2018-1}. 
However, they do not solve the problem of multiple driver set configurations. 
On the other hand, technical and experimental constraints often limit the possibility to stimulate many driver nodes in parallel, for example in gene expression modulation \cite{lodish_gene_2000} or brain stimulation \cite{hallett_transcranial_2000}.
In these cases, approaches that focus on the ability of single driver node to control the entire network, such as control centrality \cite{liu_control_2012} or single-node controllability \cite{gu_controllability_2015}, do circumvent the multiplicity issue, but can still suffer from numerical errors and approximate results.

To overcome this impasse, we propose an integrated method that combines the advantages of the previous approaches and quantifies the capacity of a single driver node to control a predefined target set.
Based on the Kalman controllability condition, our method identifies the part of the target set that can be controlled by a candidate driver. To do so, we introduce a ranking among the target nodes and we iteratively evaluate the controllability of the system by adding one target node at a time in a descending order. This eventually finds a univocal controllable configuration corresponding to the highest ranking.
In the following, we first illustrate how our method, named \textit{step-wise target controllability}, works for simple network structures and we discuss the potential benefits for directed and sparse networks as compared to alternative approaches.
Then, we use it to study molecular networks of macrophage pro-inflammatory activation, derived from ontology-based reconstructions, and identify the driver-target pathway alterations using gene expression data from blood samples of patients affected by multiple sclerosis (MS) and a matched group of healthy controls (HC).
\section{Results}
\label{sec: results}


\subsection*{Step-wise target controllability identifies a controllable subset of targets}


Let $\mathcal{G}$ be a directed graph (or network) of $N$ nodes (or vertices) and $L$ links (or edges), and $\mathcal{T}$ an arbitrary subset of $S<N$ nodes in the network.
The aim is to measure the ability of each node to drive the state of the target set $\mathcal{T}$ from a dynamical system perspective \cite{sontag_mathematical_1998}. 
In the case of linear time-invariant dynamics, the number of controllable target nodes can be obtained computing the rank of the target controllability matrix \cite{murota_note_1990, gao_target_2014, commault_functional_2019-1}:

\begin{equation}
	Q_{\mathcal{T}}=\begin{bmatrix} C B & CAB & CA^{2}B & \cdots & CA^{N-1}B \end{bmatrix} 
	\label{eq: target controllability matrix}
\end{equation}

where $A$ is the adjacency matrix of the network, $B$ is a vector identifying the driver node and $C$ is a matrix selecting the rows of $A$ corresponding to the target, or output, nodes (\textbf{\nameref{sec: methods}}).

A \textit{walk} in the network consists of an alternating sequence of vertices and edges.

The $(i,j)$ entry of $Q_{\mathcal{T}}$ indicates how many walks of length $j-1$ connect the driver to the target $i$ \cite{biggs_algebraic_1974}.   
Trivially, all the nodes not traversed by these walks do not contribute to the walk lengths and they can be neglected for the purpose of control.
By removing the irrelevant nodes from the network, $A$ becomes smaller and this results in a target controllability matrix with less columns. Put differently, we avoid the computation of matrix exponentials corresponding to non-existing driver-target walks (\textbf{\nameref{sec: methods}}). 
In practice, this can be of great advantage for reducing the occurrence of round-off errors during the matrix rank calculations.
For example, this is the case for sparse and directed networks, where fewer nodes are reachable as compared to dense and undirected networks.

This can be easily appreciated in the following example. Let us consider a directed full binary tree with $h=6$ levels, with the root node as the candidate driver. Without loss of generality, we randomly position a target in each level and we rank them according to their height in the tree.
Then, we introduce a simple cycle among the first three nodes of the tree (\textbf{Fig. \ref{fig: binary tree}}). By construction, this configuration is controllable and the entire target set can be fully driven by the driver.
However, when considering the entire network the returned rank is deficient. Instead, by removing the part of the network that is irrelevant for the control, the rank is full and we retrieve the entire controllable configuration, even in the case of larger networks, i.e. up to $h=10$ levels. 

\begin{figure}[H]
    \centering
    \includegraphics[width=\textwidth]{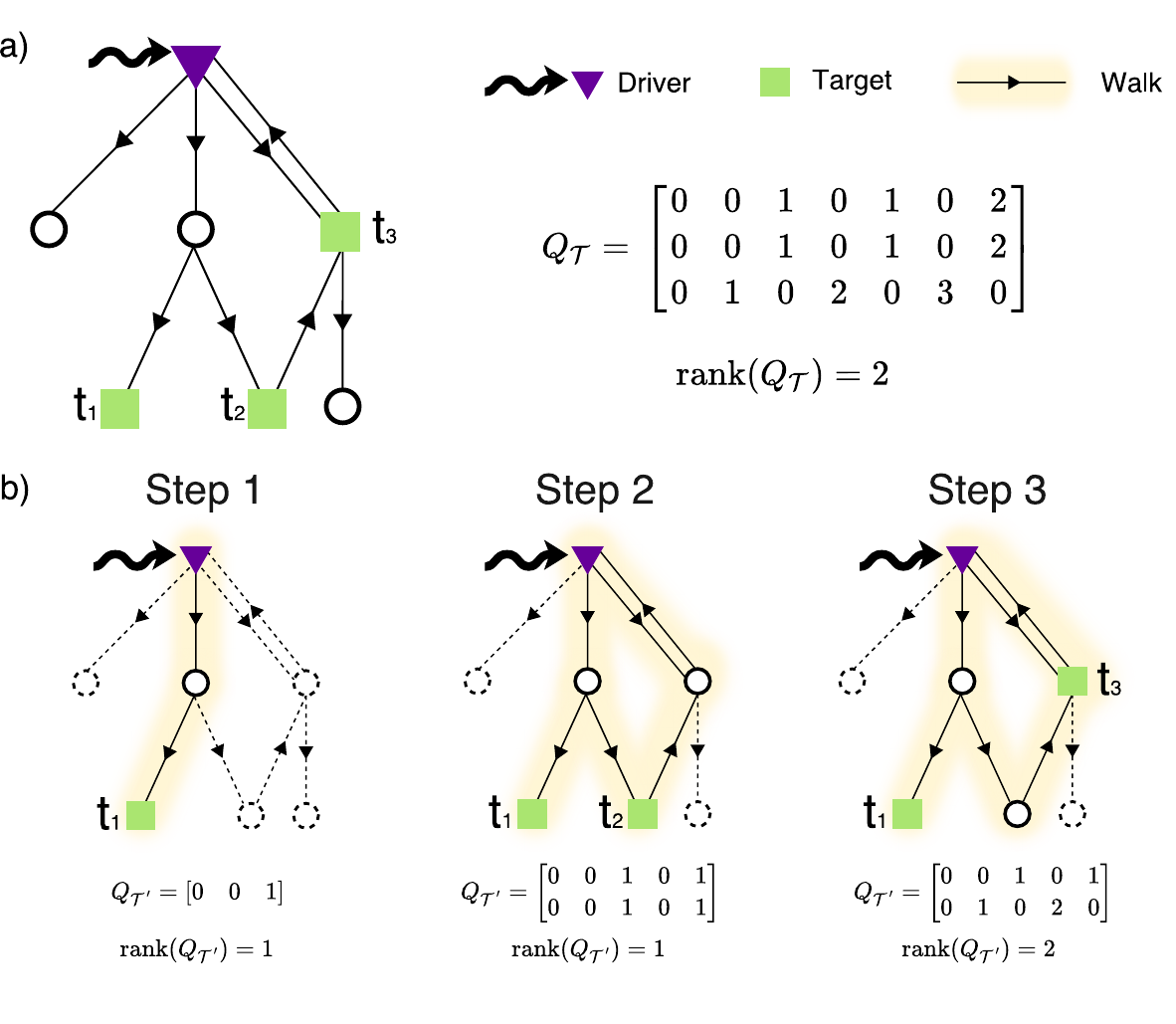}
    \caption{\small 
    \textbf{Working principle of step-wise target controllability.}
    Panel \textbf{a)} illustrates a network with one driver and a target set $\mathcal{T}=\{ t_1, t_2, t_3\}$ of cardinality  $S=3$.
    The Kalman condition informs us that only two targets are controllable from the driver, i.e. $\tau = \operatorname {rank} (Q_{\mathcal{T}}) = 2$. However, there might be up to $3$ equivalent configurations that are controllable, i.e. $\{ t_1, t_2\}$, $\{ t_1, t_3\}$, and $\{ t_2, t_3\}$. For larger networks, the number of Kalman tests to perform can be prohibitive, i.e. $\binom{S}{\tau}$.
    Panel \textbf{b)}. By introducing a hierarchy among the target nodes, our step-wise method identifies the configuration with the most important nodes by performing only $S$ tests (see \textbf{\nameref{sec: methods}}). In this example, the first step considers the subgraph containing all the walks from the driver to the target set $\mathcal{T'}= \{t_1\}$. The associated controllability matrix has full rank, i.e. $\operatorname {rank}(Q_{\mathcal{T'}})=1$. The first target is therefore retained and the algorithm moves to Step $2$, by constructing a new subgraph containing the walks from the driver to the target set $\mathcal{T'} = \{ t_1, t_2 \}$. The rank of the new controllability matrix is now deficient and $t_2$ is not retained. In Step $3$, the new subgraph contains the walks from the driver to $\mathcal{T'} = \{ t_1, t_3 \}$. Because $\operatorname {rank}(Q_{\mathcal{T'}})$ is full and there are no more targets, the algorithm stops and returns the controllable configuration $t_1,t_3$.
    }
    \label{fig: toy-model}
\end{figure}

The rank of $Q_{\mathcal{T}}$ gives the number of target nodes $\tau \leq S$ controllable by the driver, but there might be in general many possible equivalent configurations. 
To overcome this issue, we propose a step-wise procedure that tests the controllability on subproblems of increasing size.
First, we introduce a hierarchy among the target nodes and relabel them according to their importance in a descending order, i.e. $t_1 \succ t_2 \succ ... \succ t_{S}$.
Then, we create an empty auxiliary set $\mathcal{T'}$ and we sequentially include the target nodes according to their ordering. At each step, if the rank of $Q_{\mathcal{T'}}$ is full, the new target node is retained, otherwise it is removed from $\mathcal{T'}$. When all the target nodes have been visited, the algorithm returns the set of controllable targets with highest ranking (\textbf{Fig. \ref{fig: toy-model}}, \textbf{\nameref{sec: methods}}).

Our method, named \textit{step-wise target controllability}, not only returns for each candidate driver the number of controllable target nodes $\tau$ corresponding to the configuration with highest ranking, but also the set $\mathcal{T}'$ of controllable targets.

To explore the limits of our method in terms of computational complexity, we performed a simulation analysis on synthetic random networks, which vary in number of nodes, connection density and target size (\textbf{Fig. \ref{fig: computational-complexity}}). 
%
Results show that in general our method is able to retrieve a larger number of controllable targets as compared to a standard approach that computes the rank of the full controllability matrix. More specifically, when the target set contains $5\%$ of the network nodes, results are quite stable across different connection densities. For larger target-set sizes our method works better when the connection density is relatively low ($0.02$-$0.10$).
It is important also to notice that the computation of the rank starts to fail in correspondence of larger and denser networks (i.e., $N > 180$ and density $> 0.20$).


\subsection*{Driver genes are homogeneously distributed in the macrophage network}

\begin{figure}[H] 
	\includegraphics[width=\textwidth]{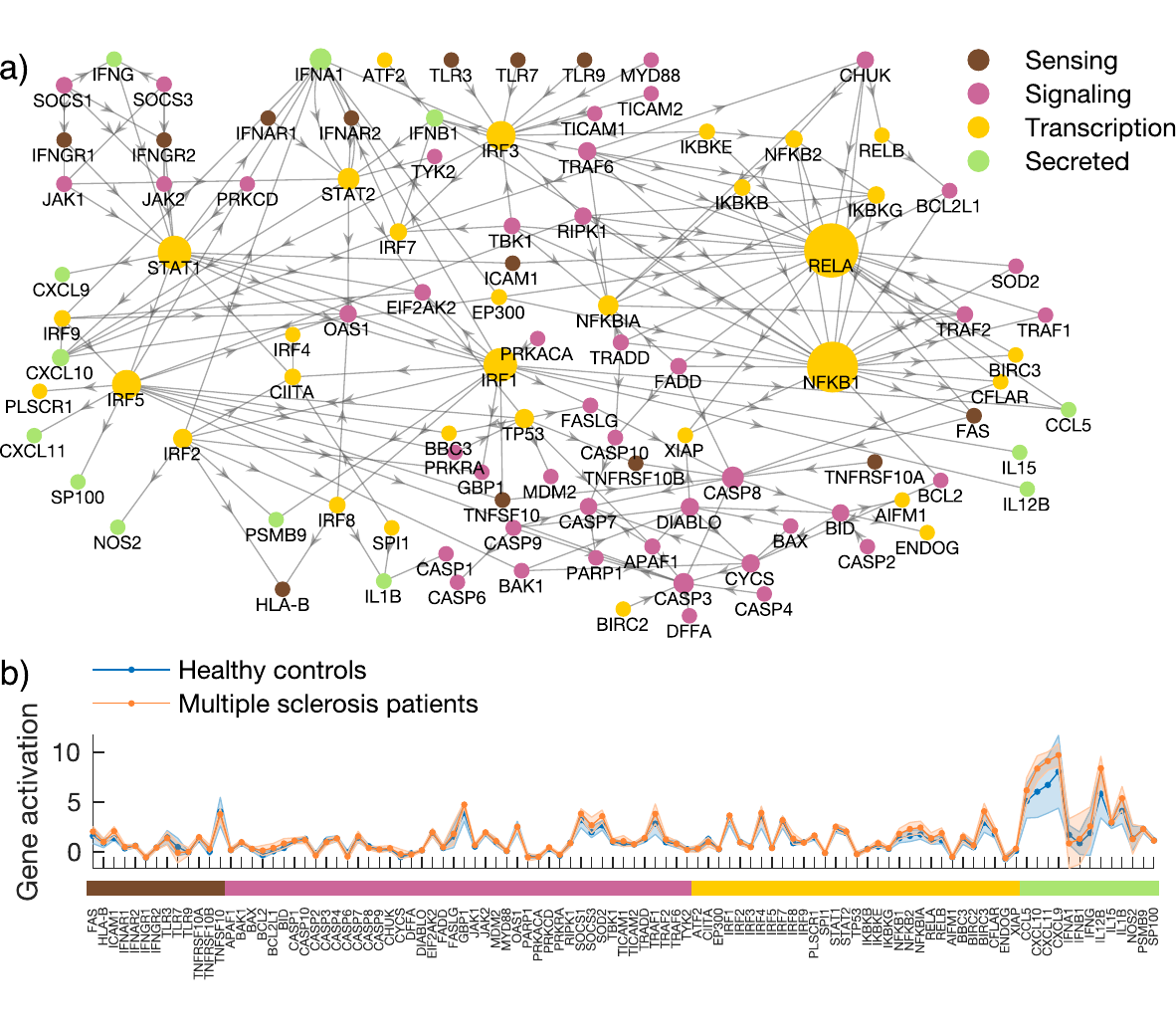}
	\centering
	\caption{\small \textbf{Molecular network and gene activation associated with the pro-inflammatory state of macrophages}.
	Panel \textbf{a)} shows the molecular network reconstructed through ontology-based techniques from the \url{macrophage.com} repository \cite{robert_macrophages.com_2011, raza_logic-based_2008}. The network consists of $N=101$ nodes corresponding to genes involved in inflammation; for the sake of interpretablity, they are organized in four classes, depending on their function in the cell. 
	\textit{Sensing} genes are in the membrane of the cell and start a \textit{signaling} pathway inside the cell, to the \textit{transcription} factors, which promote the production of \textit{secreted} molecules. 
	There are $L=211$ directed edges representing either activation or inhibition interactions between molecules (\textbf{\nameref{sec: methods}}). The size of the nodes is proportional to their total degree $k$.
    Panel \textbf{b)} shows gene activation computed as the ratio in expression between the ``pro-inflammatory'' and ``alert'' states, based on our RNA sequencing data, generated from monocyte-derived macrophages from blood samples of multiple sclerosis patients ($n=8$) and healthy controls ($n=8$) (\textbf{\nameref{sec: methods}}). Solid lines represent group-averaged values, while transparent patches stand for standard deviation.
	}
	\label{fig: network-statistics}
\end{figure}

To test our method in a biological context, we construct a network representing the interactions between molecules involved in macrophages response to pro-inflammatory stimuli (\textbf{Fig. \ref{fig: network-statistics}}), with the connections between genes inferred from a previously established network based on literature \cite{robert_macrophages.com_2011, raza_logic-based_2008}. This network is of interest in MS due to the chronic inflammation characteristic of the disease, and the generally destructive effects of pro-inflammatory macrophages in MS \cite{bitsch_tumour_2000}. Hence, dysregulation of macrophages may lead to aggravated inflammation and disease.

In order to facilitate biological interpretation of the network, we divide the nodes according to molecular function: \textit{sensing}, \textit{signaling}, \textit{transcription} factors or \textit{secreted} molecules (\href{Supplementary/Tab-S1-genes-classes.xlsx}{\textbf{Tab. S1}}). We choose the 13 \textit{secreted} molecules as target nodes because they represent the end-products of macrophage pro-inflammatory activation and enable propagation of inflammation to other cells, thus exacerbating chronic inflammation.

To establish a hierarchy among the targets, we use macrophage RNA expression data from a group of MS patients and healthy controls. The macrophages were tested with and without activating stimuli to mimic the pro-inflammatory response. We measure the gene activation as the ratio of the expression between the ``pro-inflammatory” and ``alert” condition.
We then consider the fold change $\Delta$ between the gene activation of MS patients and HC subjects (\textbf{\nameref{sec: methods}}). Genes with larger $\Delta$ values are ranked first (\textbf{Fig. \ref{fig: ordered targets}}).

Results show that $51\%$ of the tested network nodes can control at least one target (i.e. $\tau>0$) and that those drivers tend to be homogeneously distributed across classes (\href{Supplementary/Tab-S2-drivers-targets.xlsx}{\textbf{Tab. S2}}). This indicates a high redundancy in the way the target set can be controlled. Notably, target centrality values are weakly correlated (Spearman rho $0.18$, p $<0.07$) with the corresponding total node degree $k$, as defined in \textbf{\nameref{sec: methods}}, indicating that the most connected genes (e.g. RELA, NFKB1) are not necessarily the ones that can most efficiently steer the state of the target set (\textbf{Fig. \ref{fig: robustness}a}).

Almost all of the driver nodes identified by our method can control the target genes CCL5, CXCL10, CXCL11, IFNA1 and IFNB1, which code for inflammatory chemokines and cytokines. This implies a high level of co-regulation among these molecules, with many different actors exerting control over this regulation.
Interestingly, the drivers with the highest target control centrality values (SOCS1 and SOCS3, $\tau=10$) belong to feedback systems that control pro- and anti-inflammatory signal transduction by regulating the signaling process triggered in response to IFN$\gamma$ \cite{mccormick_regulation_2015}. In addition, all drivers with $\tau\geqslant9$ can be seen in our network as a cluster of genes converging onto and including STAT1 (\textbf{Fig. \ref{fig: robustness}a}). This cluster includes the receptors of IFN$\gamma$ and the signaling molecules responsible for their intracellular effects. This result matches the well-described effects of IFN$\gamma$ on chemokine production \cite{koper_cxcl9_2018,liu_interferon_2005} and overall macrophage activation \cite{mosser_exploring_2008}.

\subsubsection*{Robustness of driver nodes to random attacks}
\begin{figure}[H]
    
    \includegraphics[width=\textwidth]{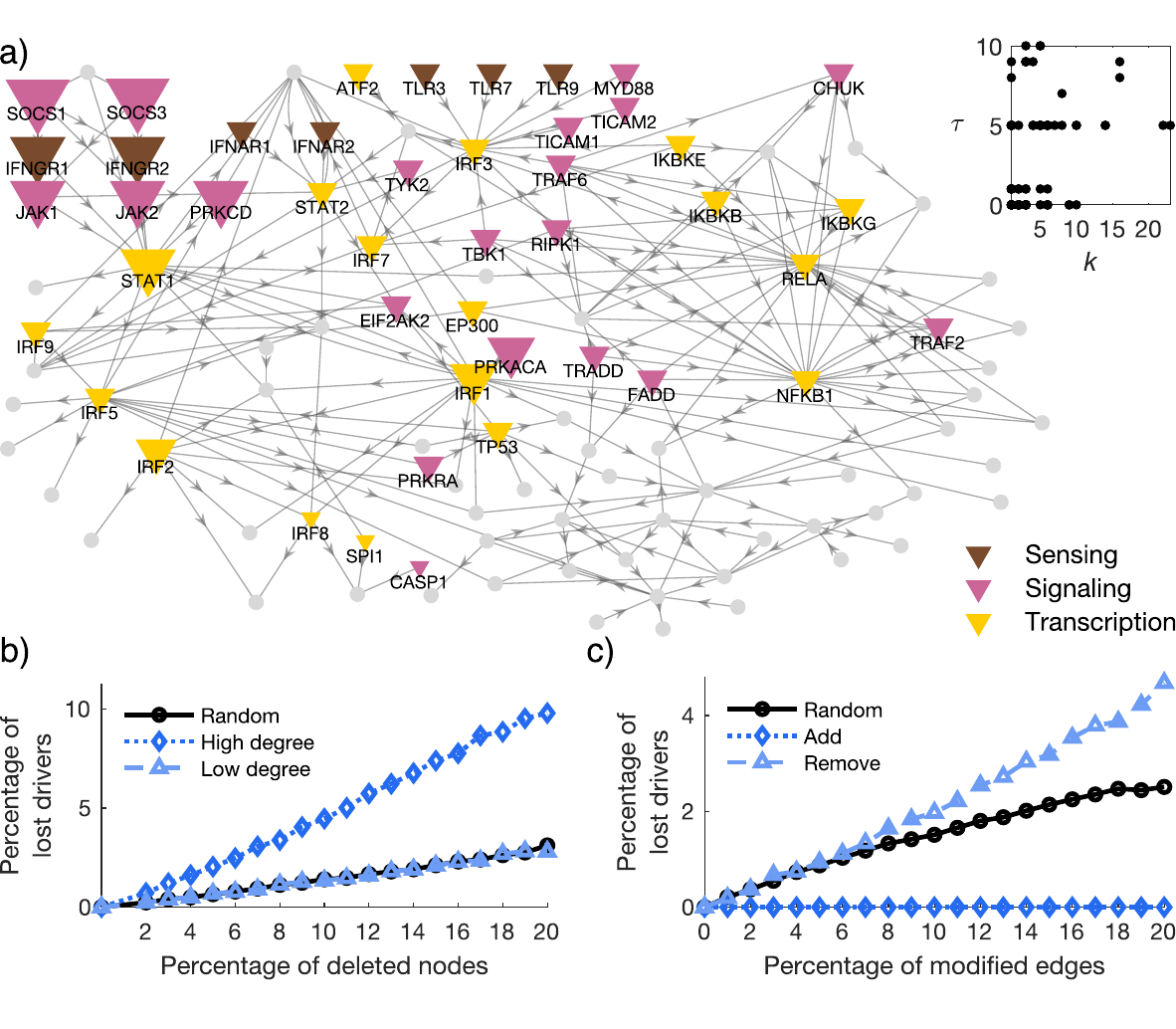}
\centering
\caption{\small 
\textbf{Gene network target control centrality and analysis of robustness for the driver nodes.}
In panel \textbf{a)} the size of the nodes codes the step-wise target control centrality values $\tau$. Nodes with $\tau=0$ are classified as not-drivers and are represented in gray. The inset shows that $\tau$ values cannot be merely predicted by node degree $k$ (Spearman rho $0.18$, p $<0.07$).
Panel \textbf{b)} shows the percentage of driver nodes ($\tau>0$) that are lost when removing nodes in a random fashion (black circles), or preferentially attacking high-degree (blue diamonds) or low-degree nodes (light blue triangles).  
Panel \textbf{c)} shows the percentage of driver nodes that are lost when randomly rewiring (black circles), adding (blue diamonds) or removing edges (light blue triangles).  
	}
	\label{fig: robustness}
\end{figure}

To assess the stability of our findings to possible errors in the network construction, we performed a robustness analysis simulating different types of alterations to its nodes and links (\textbf{\nameref{sec: methods}}).
Results show that removing nodes with higher degree $k$ leads to a greater reduction of control centrality in the drivers compared to the removal of low-degree nodes or random removal of nodes (\textbf{Fig. \ref{fig: robustness}b}). For example, by attacking $10\%$ of the nodes we lose $5\%$ of the drivers in the latter cases, while we lose $20\%$ of the drivers when removing the most connected ones (\textbf{Fig. \ref{fig: robustness}b}). 
This result confirms the crucial role of hubs in biological networks in terms of resilience to random attacks \cite{barabasi_network_2004} and controllability \cite{pu_robustness_2012}.

When perturbing links, the worst condition is given by their random removal. By attacking $10\%$ of the links around $5\%$ of the drivers are lost. This is intuitively due to the interruption of driver-target walks and to consequent impossibility to control a node that cannot be reached.
While randomly rewiring the links has an intermediate impact, adding new links has no effect on the target control centrality of the drivers (\textbf{Fig. \ref{fig: robustness}c)}.
This is of great advantage as it shows that our results will not change if new connections are established or provided by the literature.


\subsection*{Gene dysregulation and altered driver-target coactivation in multiple sclerosis}

\begin{figure}[H] 
	\includegraphics[width=\textwidth]{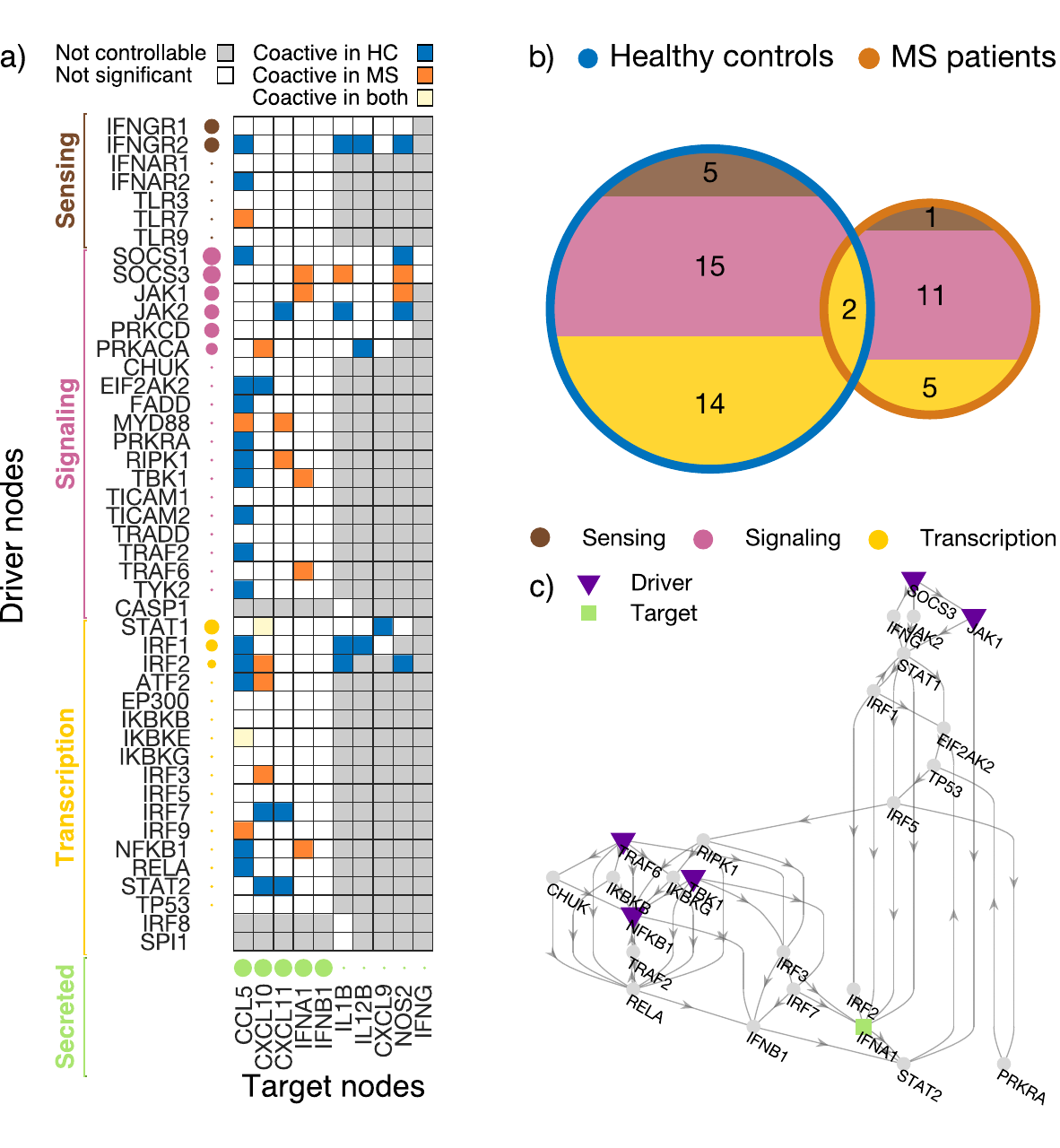}
	\centering
	\caption{\small 
	\textbf{Altered driver-target coactivation in multiple sclerosis.}  
	Panel \textbf{a)} reports the coactive driver-target pairs, computed as significant Spearman correlations (p$< 0.05$) between the gene activation of controllable driver-target pairs, for the healthy control (HC - blue squares) and the multiple sclerosis group (MS - red squares). 
	White squares indicate that there is a controllable walk from the driver to the target, but that their correlation is not significant. Grey squares mean that there is no controllable walk for driver-target pairs.
	The size of the circles for driver nodes codes for their target control centrality values $\tau$. For target genes, circle sizes represent the number of driver nodes that can control them.
	Panel \textbf{b)} Venn diagram showing a decrease in number of driver-target coactivations in the MS patients as compared to HC. In both groups, these functional interactions tend to predominantly involve signaling genes. 
	Panel \textbf{c)} subnetwork of the walks from all the drivers coactivated with the target IFNA1. 
	}
	\label{fig: corr}
\end{figure}

Using step-wise target controllability, we detect potential directed interactions in the macrophage activation network, but we cannot quantify how changes in the driver's state affect those in the targets.
To measure driver-target functional interactions, we compute the Spearman correlation between the gene activation of controllable driver-target pairs, for the HC and MS groups  (\textbf{\nameref{sec: methods}, \href{Supplementary/Table-S3-correlations_pvalues}{\textbf{Tab. S3}}}). 
We call \textit{coactivated} the genes exhibiting a significant correlation (p$<0.05$).
Results show that in general only a moderate fraction ($21\%$) of all the possible driver-target genes are coactivated (\textbf{Fig. \ref{fig: corr}a, \href{Supplementary/Table-S3-correlations_pvalues}{\textbf{Tab. S3}}}). For both HC and MS groups these interactions tend to primarily involve signaling functions \textbf{Fig. \ref{fig: corr}b}.
However, the number of driver-target coactivations is lower in the pathological condition ($MS=19$ versus $HC=36$). More importantly, they differ from those observed in the HC group (\textbf{Fig. \ref{fig: corr}b)}. 
This is particularly evident for target IFNA1, which only exhibits coactivations with signaling and transcription drivers in the MS group (\textbf{Fig. \ref{fig: corr}c}).

\begin{figure}[H] 
	\includegraphics[width=\textwidth]{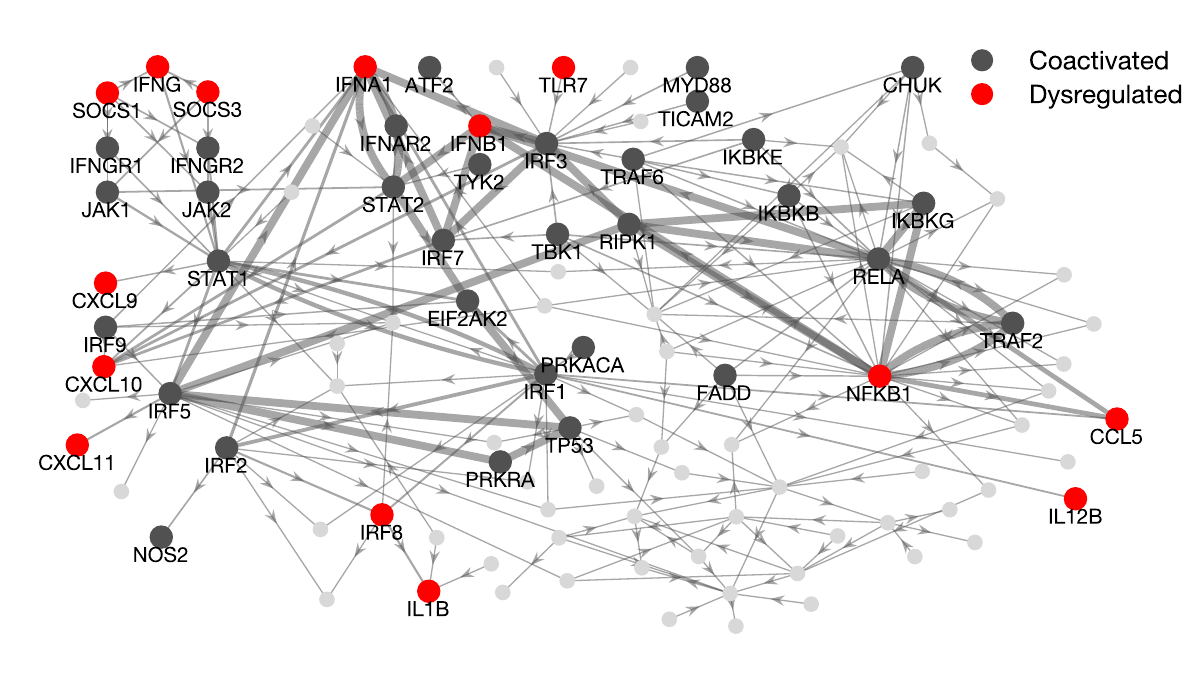}
	\centering
	\caption{\small \textbf{Pooled visualization of dysregulated genes along differentially coactivated driver-target walks.} Highlighted genes indicate all nodes on walks between coactivated driver-target pairs, either in the healthy control group, or in the MS patients group.
	Dysregulated genes are shown in red. Edge thickness is proportional to the number of times they are traversed by walks connecting a driver to a target node (information not reported here).}
	\label{fig: interpretation}
\end{figure}

Because the macrophage network edges are fixed and reconstructed from known protein-protein interactions, differences in coactivation can be essentially attributed to altered regulation of transcription.
Hence, our hypothesis is that the observed functional reorganization can be explained by the dysregulation of specific genes along the controllable walks from the drivers to targets.
To test this prediction, we examine all the pairs of genes whose coactivation appears or disappears in the MS group (\textbf{Fig. \ref{fig: corr}a}).
We found that $47/51$ of these differentially coactivated pairs present at least one dysregulated gene (i.e., fold change $\lvert \Delta \lvert$ above the $75$-percentile, \textbf{\nameref{sec: methods}}) on the walk from the driver to the target (\textbf{Fig. \ref{fig: interpretation}, \href{Supplementary/Table-S4-dysregulated-genes}{\textbf{Tab. S4}}}).
%


We find in total $14$ dysregulated nodes on any of these walks. The genes that most frequently appear are NFKB1, IFNA1 and IFNB1 ($36/51$ walks). 
They are present on all walks that end with targets CCL5, CXCL10, CXCL11, IFNA1 or IFNB1, i.e., the 5 targets that could be controlled by most drivers. This points to their dysregulation being a potent disruptor of the normal network functioning.
The co-occurrence of these three dysregulated genes can be explained by a feedback loop in which NFKB1 activates IFNB1, and IFNA1 and IFBN1 both activate STAT2, which through several intermediates can influence all three genes (\textbf{Fig. \ref{fig: connected-component}}). Indeed, this stems from the fact that all these nodes belong to the main connected component of the network, i.e. a subnetwork in which every node is reachable from any other node.
%

Taken together, these results indicate that the aberrant reorganization of functional interactions in the MS group is associated with the presence of dysregulated genes along the controllable walks of the macrophage network.

\subsubsection*{Switch of SOCS-gene coactive drivers reflects dysregulated inflammatory response}

\begin{figure}[H] 
	\includegraphics[width=\textwidth]{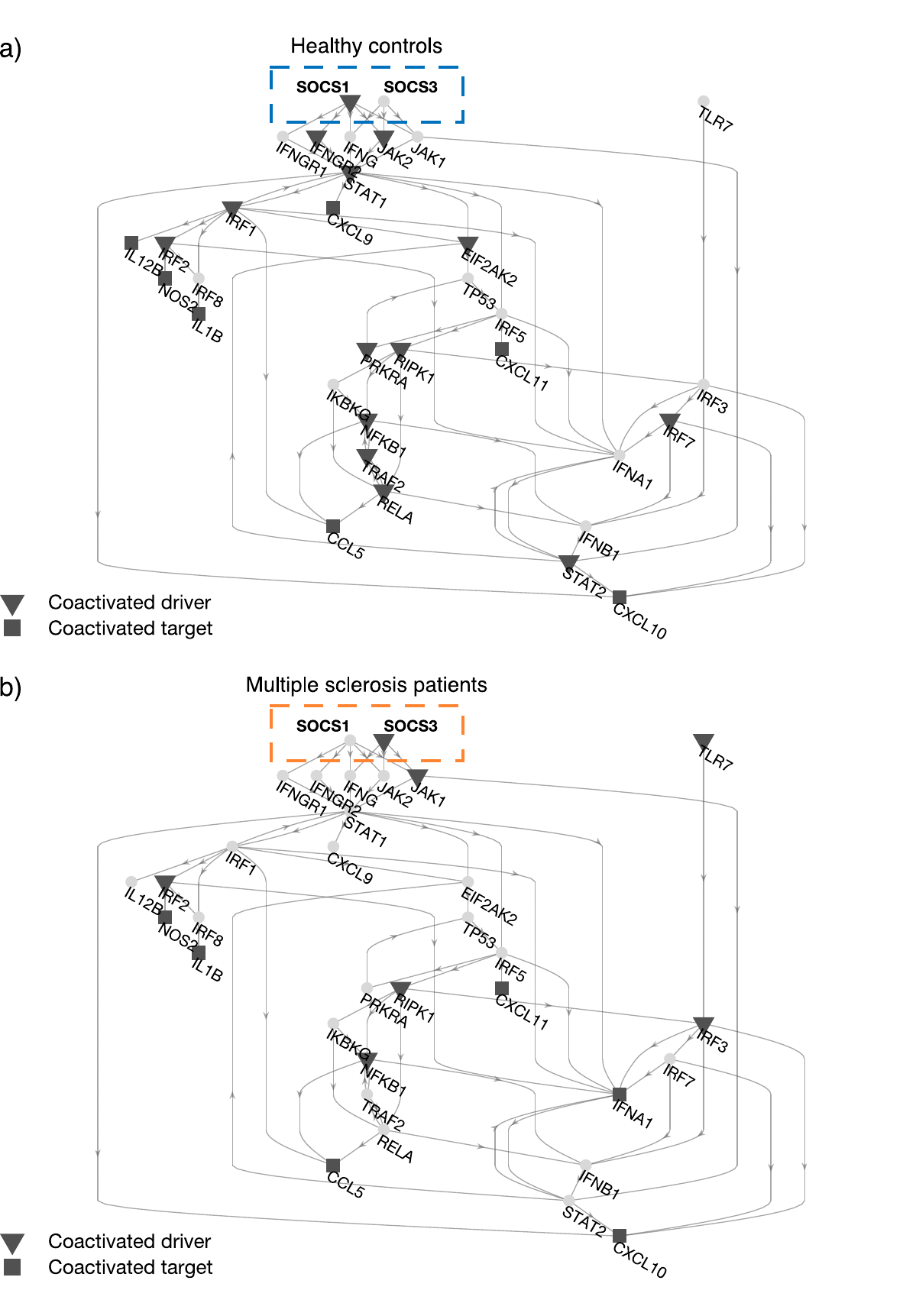}
	\centering
	\caption{\small \textbf{Dyregulated drivers and coactivation switch for SOCS-genes.} The subnetwork includes all dysregulated drivers (IRF8, NFKB1, SOCS1, SOCS3, TLR7) and their controllable targets.
	Panel \textbf{a)} shows coactivated pairs for healthy controls (HC), panel \textbf{b)} shows coactivated pairs for multiple sclerosis (MS) patients. A coactivation switch can be appreciated between the HC and MS group. SOCS1 and SOCS3 are respectively coactive and silent in HC, while they invert their role in the MS group.
	}
	\label{fig: bio-interpretation-dys}
\end{figure}

Because drivers are crucial for steering the target network's state, we focus on the subnetwork specifically involving the dysregulated drivers (IRF8, NFKB1, SOCS1, SOCS3, TLR7) and the walks towards the respective controllable targets \textbf{Fig. \ref{fig: bio-interpretation-dys}}. 
By looking at how driver and target nodes are differently coactivated in healthy controls and MS patients, we obtain a much clearer description of the gene dysregulation effects.
First, many of the previous results can be now appreciated in finer detail, such as i) the reduction in number of coactivated driver-target pairs in MS, ii) the large number of targets that can be controlled by SOCS1 and SOCS3, and iii) the potential of NFKB1, IFNA1 and IFNB1 to affect the driver-target functional interactions. 

Second, we report an interesting mechanism involving the drivers with the highest $\tau$ centrality values, i.e. SOCS1 and SOCS3.
In the HC group, SOCS1 is coactivated with the targets while SOCS3 does not exhibit any significant correlation. 
In the MS group, we observe the opposite, i.e. SOCS1 is silent while SOCS3 becomes coactive.
Because both driver genes are dysregulated, the observed ``switch'' mechanism could be therefore associated with the altered pro-inflammatory response of the MS group. 
Indeed, these two molecules are known to be strong modulators of macrophage response: SOCS1 inhibits the signaling of pro-inflammatory genes while SOCS3 is known to be an important actor in inflammatory response, with the ratio of the two proteins determining the actual effect \cite{wilson_socs_2014}.

\section{Discussion}
\label{sec: discussion}



\subsection*{Identification of controllable configurations in complex networks}

Network controllability refers to the ability to drive an interconnected dynamical system from any initial state to any desired final state in finite time, through a suitable selection of inputs \cite{rugh_linear_1995, sontag_mathematical_1998}.
In recent years, an increasing number of research groups from different disciplines have focused their efforts on identifying the minimum set of driver nodes or quantifying the capacity of single nodes to control the entire network, as well as parts of it \cite{liu_control_2016, lugagne_balancing_2017, sun_controllability_2016, wuchty_controllability_2014, tang_developmental_2017, dosenbach_distinct_2007, wagner_noninvasive_2007, menara_structural_2017, gu_optimal_2017, betzel_optimally_2016, muldoon_stimulation-based_2016}. 
Despite being theoretically attractive, network controllability still suffers from computational issues that limit its impact in concrete applications. This is mainly due to the presence of multiple equivalent controllable walks in a network that make the associated controllability problem ill-posed and/or the resulting solution space very big \cite{rugh_linear_1995, sontag_mathematical_1998}.

To reduce such complexity, we propose a method based on control centrality, which was previously designed to quantify the ability of one node to control directed networks \cite{liu_control_2012}.
First, we define the \textit{target} control centrality to measure the controllability of a specific part of the network, i.e., a predefined target set.
Because edges are directed, this has the advantage to ignore the part of the network that is not traversed by the walks connecting the driver to the target set.
Second, we introduce an ordering among the target nodes and perform a step-wise controllability test with increasing size. 
Because of the ranking, only one controllable configuration will be identified, i.e. the one with the highest ranking  (\textbf{Fig. \ref{fig: toy-model}}).
To test the controllability of the driver-targets configuration at each step we adopted the Kalman criterion \cite{kalman_mathematical_1963, rugh_linear_1995}. However, the entire iterative framework is quite flexible and other methods, such as Gramian condition \cite{klickstein_energy_2017}, Popov-Belevich-Hautus criterion \cite{li_structural_2018-1}, or feedback vertex set \cite{zanudo_structure-based_2017}, could be used as alternative controllability criteria.

The main advantage in terms of time of our method consists in the fact that we identify a controllable driver-targets configuration by performing $j$ tests when $j<S$ out of $S$ target nodes are controllable, as opposed to a brute-force approach that requires to test all the possible configurations with $j$ targets and that, in the worst-case scenario, leads to ${j}\choose{S}$ tests.

In addition, because of the step-wise procedure, the method does not need to compute the rank of the full controllability matrix and this allows to minimize possible round-off errors at least for relatively small networks. Specifically, our results show that the computations are reliable when $N<180$ and the connection density is $< 0.2$ (\textbf{Fig. \ref{fig: computational-complexity}}).
While these numbers are in general relatively low and impede to scale up to very large networks, they still represent ranges that are compatible with typical brain networks obtained from neuroimaging data (\cite{bullmore_complex_2009, bullmore_brain_2011, de_vico_fallani_graph_2014}).


\subsection*{Control pathways in macrophage molecular networks}
The study of the molecular interactions is crucial to the understanding of the basic functions of the cell such as proliferation or apoptosis \cite{taylor_dynamic_2009,maslov_specificity_2002}. 
Determining the connection mechanisms that rule a specific biological function can significantly impact our daily life by providing new therapeutics to counteract diseases \cite{drier_pathway-based_2013, menche_disease_2015, tong_novel_2017}.
Studying molecular networks is however difficult, because in general we do not know the true functional interactions of a cell and indirect techniques such as gene co-expression are typically employed to infer such connections \cite{uygun_utility_2016}. Based on correlation analysis, these methods cannot inform on the causal nature of the interactions. More importantly, the reliability of the estimated network critically depends on the number of interactions to number of data samples ratio, which is in practice very low \cite{uygun_utility_2016}.

To overcome this limitation, we reconstruct the directed gene interactions associated with the inflammatory state of the human macrophages by adopting a novel ontology-based approach that integrates the available information from multiple datasets and results in the literature \cite{henry_bipom_2017}. 
Previous studies show that the number of driver nodes in biological networks is rather high due to their sparse and heterogeneous nature \cite{liu_controllability_2011, ruths_control_2014}.
Consistently, we find that a large percentage of genes ($51\%$) can control at least one secreted molecule in the target set. Our results also confirm that, despite being crucial for global communication, hubs (e.g. RELA) are not always the most important from a network control perspective (\textbf{Fig. \ref{fig: robustness}}a). This stems from the theoretical impossibility to diversify the input signals to all the connected neighbors \cite{liu_controllability_2011}. 
 The found driver genes are heterogeneously distributed across the tested gene classes.
However, our method highlights SOCS1 and SOCS3 as the drivers with the highest target control centrality values, with other IFN$\gamma$-response-related genes showing similar values. This is in line with the known effects of SOCS-genes and IFN$\gamma$ on molecules secreted by pro-inflammatory macrophages \cite{mccormick_regulation_2015, koper_cxcl9_2018, liu_interferon_2005, mosser_exploring_2008}, supporting the ability of this method to identify biologically relevant drivers.

Overall, these results uncover the existence of potential causal influences from candidate driver genes to the secreted molecules in the human macrophage activation network. 
Because the identified driver nodes are robust to network alterations, notably when adding new links (\textbf{Fig. \ref{fig: robustness}}cblue diamonds), the obtained results are expected to be sufficiently resilient to the integration of new gene-gene interactions.
Notably, our results are also relatively robust to the deletion of links, a situation which is often associated with filtering or thresholding out the weakest or less relevant connections in biological networks \cite{fallani_topological_2017}
(\textbf{Fig. \ref{fig: robustness}c} light blue triangles).

From a different angle, our approach can be seen as a principled manner to focus on specific nodes (the drivers) or node pairs (driver-target) in complex networks. This might have important consequences when studying genome-wide databases where the high number of elements can make prohibitive the assessment of significant gene expressions and/or co-expressions \cite{bentley_accurate_2008, mullighan_genome-wide_2007}.

\subsection*{Dysregulated genes and aberrant interactions in multiple sclerosis}

Multiple sclerosis is an immune-mediated disease in which the immune system erroneously attacks myelin in the central nervous system. There are many neurological symptoms, including motor and cognitive deficits, that can vary in type and severity depending on the attacked central nervous system regions \cite{hauser_multiple_2015}. 
The role of macrophages in MS is crucial because of their ability to obtain a pro-inflammatory activation state, including the release of pro-inflammatory cytokines and leading to central nervous system tissue damage \cite{chu_roles_2018}. Hence, dysregulation of macrophages may lead to autoimmunity and persistent inflammatory diseases \cite{strauss_immunophenotype_2015}. While the etiology of MS is still not well-understood there is a large consensus on its genetic basis and on the importance of unveiling the underlying network mechanisms \cite{airas_hormonal_2015}. 

In this study, we combined network controllability tools and gene expression data to detect the genes responsible for altering the macrophages action in multiple sclerosis.
Differently from standard approaches, where the attention is focused on the identification of the driver nodes in a network, we here propose an alternative way of exploiting network controllability.
We first show that the macrophage inflammatory state in the MS group was characterized by a drastic alteration of the coactivations in the driver and target genes (\textbf{Fig. \ref{fig: corr}}).
Such absence of coordination was in general associated with the presence of dysregulated genes along the walks from the driver to the target node.
Notably, the pathological dysregulation of NFKB1, IFNB1 and IFNA1, which belong to the same feedback cycle (\textbf{Fig. \ref{fig: connected-component}}), critically affects several driver-target functional interactions (\textbf{Fig. \ref{fig: interpretation}}).

Finally, our approach allows to identify a shift mechanism for dysregulated SOCS1 and SOCS3 drivers, showing opposite coactivation patterns in MS patients compared to the healthy controls (\textbf{Fig. \ref{fig: bio-interpretation-dys}}).
These results suggest that experimentally stimulating SOCS3 - a strong inducer of pro-inflammatory response - might be more effective for moving the state of the altered secreted molecules towards physiological configurations.
Taken together, these results might have practical consequences on how to design intervention strategies and counteract disease phenotype.


\subsection*{Methodological considerations}
Our method uses Kalman controllability rank condition \cite{kalman_mathematical_1963} to quantify the centrality of the driver nodes. This criterion assumes that the investigated system has a linear dynamics, \textbf{Eq. \ref{eq: LTI sys}}.
In our case, this means that the changes in the gene activation would follow a linear trend. 
While this is in general not true and difficult to ascertain, it appears that results from non-linear tests are often dominated by linear relationships \cite{song_comparison_2012, steuer_mutual_2002}. 
Furthermore, a significant fraction of the data analysis and modeling deals exclusively with linear approaches as they are simpler, easy to interpret and serve as a prerequisite of nonlinear behavior \cite{liu_control_2016}.

Another peculiarity of our approach is the assumption of time-invariant interactions in the molecular gene network. 
On the one hand, this assumption allows to better exploit the well-established results and tools in network controllability \cite{liu_controllability_2011}; on the other hand, it might conflict with existing literature looking for biological connectivity changes between conditions or populations such as differential gene coexpression \cite{choi_differential_2005}.
Here, we hypothesized that the activation state of each node (in terms of gene expression) could eventually change but not the underlying network structure.
Thus, our network - obtained from detailed maps of the macrophage cells - would only act as a substrate/proxy for functional interactions, such as correlated gene activities.


Our method requires a specific ordering of the target nodes. While this can be typically achieved in many biological applications - by ranking nodes according to their state (e.g., gene expression \cite{zhao_ranking_2011, liseron-monfils_necorr_2018}, brain activity \cite{lohmann_eigenvector_2010, sen_ranking_2019}) - challenges might still remain in general.
When it is not possible to impose a ranking of the nodes from external knowledge, another possibility is to derive it from the network structure taking into account, for example, node centrality values. However, there might be multiple nodes with the same centrality value, which would impede a proper ranking. In those situations, multiple node centrality measures could be integrated to get a more heterogeneous distribution (e.g., degree, strength, betweenness, closeness) \cite{newman_networks:_2010}. Another possibility is to add equally important targets at the same iteration step and test if they are simultaneously controllable. While this procedure is suboptimal and may underestimate the number of controllable targets, it still minimizes the computational complexity related to testing multiple driver-target combinations.


We finally notice that our method is conceived for directed networks only, where the dimensionality reduction has a real computational benefit.
In fact, in the case of undirected graphs, it is not possible to remove nodes on the walks from the driver to the targets since information is bound to span the entire network. Similarly, for directed but dense networks, the possibility to focus on specific parts of the network, and reduce the computational cost, becomes lower regardless of the topology.

To conclude, it is important to mention that extensions of network controllability tools to time-varying frameworks do exist \cite{li_fundamental_2016, zhang_controllability_2017}.
However, in that case networks would be inferred from gene coexpression and therefore affected by statistical uncertainty due to sample sizes. 
Further research is needed to seek how to apply network controllability in presence of noisy time-varying connections.


\subsection*{Conclusion}
In this study, We introduce a method to quantify the ability of candidate driver nodes to drive the state of a target set within a sparse and directed network.
Further, we illustrate how this method works for the  molecular network associated with the human macrophage inflammatory response.  The obtained results reveal in a principled way the genes that are significantly dysregulated in multiple sclerosis.
We hope that this method can contribute to the identification of the key nodes in biological networks to better identify pharmacological targets to counteract human diseases.

\section{Material and methods}
\label{sec: methods}


\subsection*{Step-wise target controllability}

We introduce a method to identify which target nodes in a network that can be controlled from a single \textit{driver} node.
To do so, we start by considering the canonical linear time-invariant dynamics on a directed network described by the adjacency matrix $A \in \mathbb{R}^{N \times N}$

\begin{equation}
	\dot{\textbf{x}}(n) = A \textbf{x}(n) + B \textbf{u}(n), \qquad \textbf{y}(n) = C\textbf{x}(n)
	\label{eq: LTI sys}
\end{equation}

where $\textbf{x}(n) \in \mathbb{R}^N$ describes the state of each node at time $t$, $B \in \mathbb{R}^N$ specifies the driver node, $\textbf{u}(n) \in \mathbb{R}^N$ is its external input (or control) signal, $\textbf{y}(n) \in \mathbb{R}^S$ is the output vector, and $C \in \mathbb{R}^{S \times N}$ is the output matrix identifying the target nodes. 

Such a system is controllable if it can be guided from any initial state to any desired final state in finite time, with a suitable choice of input. 
A necessary and sufficient condition to assess the controllability of \textbf{Eq. \ref{eq: LTI sys}}, is that the controllability matrix $Q$

\begin{equation}
	Q = \begin{bmatrix} B&AB&A^{2}B&\cdots &A^{N-1}B \end{bmatrix} 
	\label{eq: kalman matrix}
\end{equation}

has full row rank, i.e. rank$(Q)=N$. That is the Kalman rank condition, which basically verifies the existence of linearly independent rows in $Q$ \cite{kalman_mathematical_1963, rugh_linear_1995}. If so, the driver node can reach and control the dynamics of all the other nodes through independent walks of length $N-1$ at maximum.

If it is of interest to control only a target set $\mathcal{T}$ of the network, specified in $C$ and consisting of $S \le N$ nodes, then \textbf{Eq. \ref{eq: LTI sys}} can be reduced into a target controllability matrix $Q_\mathcal{T} = C Q$ (\textbf{Eq. \ref{eq: target controllability matrix}}), where $C$ filters the rows of interest corresponding to the targets. 
Now, the rank of $Q_\mathcal{T}$ gives the number $\tau \leq S$ of nodes in the target set that can be controlled by the driver.

To identify a driver-target configuration, we further introduce a hierarchy among the target nodes, so that we can order and relabel them from the most important one to the least,  i.e. $t_1 \succ t_2 \succ ... \succ t_S$.
Then we perform the following step-wise procedure for each candidate driver node

\begin{itemize}

	\item Step 1. \textit{Initialization}
	
	\begin{itemize}
		\item Create a temporary empty target set $\mathcal{T'} \leftarrow \lbrace \rbrace$ 
		\item Set the number of controllable targets $\tau \leftarrow 0$
	\end{itemize}
	
	\item Step 2. \textit{Repeat until termination criteria are met.} For $j \leftarrow 1,...,S$ do
	
	\begin{itemize}
		\item Add the $j$-th target node to the target set $\mathcal{T'} \leftarrow \mathcal{T'} \cup \{ t_j \}$ 
		\item Build the subgraph containing the nodes on walks from the driver to the targets in $\mathcal{T'}$
		\item Compute the rank of the target controllability matrix $Q_\mathcal{T'}$
		\item \textit{If} rank($Q_\mathcal{T'}$) is full then $\tau \leftarrow \tau + 1$ else $\mathcal{T'} \leftarrow \mathcal{T'} \setminus \{t_j \}$
		\item $j\leftarrow j+1$
	\end{itemize}
	\item Step 3. \textit{Output $\tau$ and $\mathcal{T'}$} 
\end{itemize}

Eventually, the \textit{target control centrality} $\tau$ is the number of controllable targets in $\mathcal{T}$, and the set $\mathcal{T}'$ contains the $\tau$ controllable targets with highest ranking.

Note that in general, the method may underestimate the actual number of controllable target nodes because of the occurrence of numerical errors, but not because of the step-wise procedure itself.
The Matlab code associated to the step-wise target controllability is freely available at \url{https://github.com/BCI-NET/Public}


\subsection*{Construction of the macrophage activation network}

We reconstruct the inflammatory molecular network of the human macrophage by integrating information from the macrophage signal transduction map \cite{robert_macrophages.com_2011, raza_logic-based_2008}.
This map contains a comprehensive, validated, and annotated map of signal transduction pathways of inflammatory processes in macrophages based on the current literature. 
To extract molecular interactions from this map, we used the Hermit software \cite{motik_structured_2008}, which implements automatic reasoning based on logical rules. 
We specifically used the rules implemented in the molecular network ontology to infer molecular interactions depending on the process they belong \cite{musen_protege_2015,henry_bipom_2017}.
Because we are interested in the inflammation process, we restricted our analysis to a specific subset of $101$ genes with known roles in macrophage pro-inflammatory activation, and for which their regulation in response to pro-inflammatory stimuli could be confirmed in our data set. These genes were classified according to their function in the cell: \textit{sensing}, \textit{signaling}, \textit{transcription} and \textit{secreted} (\href{Supplementary/Table-S1-genes-classes.xlsx}{\textbf{Tab. S1}}), as described in databases such as NCBI Gene \cite{agarwala_database_2018}, UniProt \cite{the_uniprot_consortium_uniprot_2019} and GeneCards \cite{stelzer_genecards_2016}. The full network was thus reduced to only include these genes and their interactions. 
Due to recent studies, we also opted to exclude two edges (from SOCS3 to IFNGR1 and to IFNGR2) to represent the involved pathways \cite{wilson_socs_2014}.

The resulting network contains $N=101$ nodes and $L=211$ unweighted directed edges representing either activation or inhibition between genes.
The total degree $k$ of each node in the network is computed by summing the number of incoming and outgoing edges:

\begin{equation}
	k_i = \sum_{j=1}^{N}A_{ij} + \sum_{j=1}^{N}A_{ji}
	\label{eq: degree}
\end{equation}
where $A_{ij}=1$ if there is an edge between the corresponding genes, and $0$ otherwise. 

\subsection*{Collection of macrophage mRNA expression data}

Collection of blood for the study was approved by the French Ethics committee and the French ministry of research (DC-2012-1535 and AC-2012-1536). Written informed consent was obtained from all study participants. All patients fulfilled diagnostic criteria for multiple sclerosis \cite{thompson_diagnosis_2018}, and individuals (multiple sclerosis patients and healthy donors) with any other inflammatory or neurological disorders were excluded from the study. Patients were included in the study only if they were not undergoing treatment.

Blood was sampled from $8$ MS patients and $8$ healthy controls in acid citrate dextrose tubes. From blood samples, peripheral blood mononuclear cells were isolated using Ficoll Paque Plus (\url{www.gelifesciences.com}) and centrifugation ($2200$ rpm, $20$ min). Cells were washed in PBS and RPMI $+10\%$ FCS. Monocytes were isolated with anti-CD14 microbeads (\url{www.miltenyibiotec.com}) and plated in $12$-well plates ($500 000 \, \text{cells}/\text{well}$) in RPMI $+10\%$ FCS and granulocyte-macrophage colony-stimulating factor ($500 \, \text{U}/\text{ml}$) to induce differentiation into macrophages. After $72$h, media was replaced with fresh media supplemented with granulocyte-macrophage colony-stimulating factor ($500 \, \text{U}/\text{ml}$) to maintain ``alert'' macrophages or IFN$\gamma$ ($200 \, \text{U}/\text{ml}$) + upLPS ($10 \, \text{ng}/\text{ml}$) to induce ``pro-inflammatory'' activation. Cells were lysed after $24$h and RNA was extracted with RNeasy Mini Kit (\url{www.qiagen.com}).

Transcriptome sequencing cDNA libraries were prepared using a stranded mRNA polyA selection (Truseq stranded mRNA kit, \url{www.illumina.com}). For each sample, we performed 60 million single-end, 75 base reads on a NextSeq 500 sequencer (\url{www.illumina.com}). RNA-Seq data analyses were performed by GenoSplice technology (\url{www.genosplice.com}). Sequencing, data quality, reads repartition (e.g., for potential ribosomal contamination), and insert size estimation are performed using FastQC \cite{andrews_fastqc_2010}, Picard-Tools (\url{http://broadinstitute.github.io/picard/}), Samtools \cite{li_sequence_2009} and rseqc \cite{wang_rseqc_2012}. Reads were mapped using STARv2.4.0 \cite{dobin_star_2013} on the hg19 Human genome assembly. Gene expression regulation study was performed \cite{noli_discordant_2015}. 
Briefly, for each gene present in the FAST DB v2018\_1 annotations, reads aligning on constitutive regions (that are not prone to alternative splicing) were counted.
Based on these read counts, normalization was performed using DESeq2 \cite{love_moderated_2014} in R (v.3.2.5) \cite{r_core_team_r_2014}.


\subsection*{Network modeling and data analysis}
In the modeling framework described by \textbf{Eq \ref{eq: LTI sys}}, matrix $A$ corresponds to the molecular network and represents the time-invariant component of the system.
The dynamic component is instead represented by the gene activation response in the healthy and diseased condition (\textbf{Fig. \ref{fig: network-statistics}b}), computed as the ratio in gene expression between the ``pro-inflammatory'' and ``alert'' condition. Specifically, $\textbf{x}(n)$ represents the gene activation.
$B$ is a vector identifying the candidate driver. 
The control signal $\textbf{u}(n)$ is out of the scope of this work.
The output vector $\textbf{y}(n)$ and the output matrix $C$ identify the target nodes.

We select the genes belonging to the \textit{secreted} molecules class (\href{Supplementary/Table-S1-genes-classes.xlsx}{\textbf{Tab. S1}}) as our target set $\mathcal{T}$.
All the nodes in the other classes are then tested separately as potential driver nodes by computing their target control centrality $\tau$.
To enhance numerical precision, the logarithmic transformation $\log{(q+1)}$ is applied to the elements of the target controllability matrix $Q_\mathcal{T}$ (\textbf{Eq. \ref{eq: target controllability matrix}}).

The hierarchy among the target nodes is established by computing the fold change $\Delta$ between the corresponding gene activation in the two groups:
\begin{equation}
	\Delta = \frac{\mu_{MS}}{\mu_{HC} }
	\label{eq: rel change}
\end{equation}
where $\mu_{MS}$ and $\mu_{HC}$ are group-averages for MS patients and healthy controls, respectively, of the gene activation.
Nodes with higher $\Delta$ absolute values are ranked first. Highly positive $\Delta$ values indicate a too strong inflammatory response (over-activation) in the MS patients with respect to the healthy controls. Highly negative $\Delta$ values indicate a too weak inflammatory response (under-activation). 
We define \textit{dysregulated} genes along the controllable driver-target walks as those for which  $\lvert \Delta \rvert $ is above the $75th$ percentile.

We perform a robustness analysis to evaluate the stability of the identified driver nodes to potential errors in the molecular network reconstruction.
We simulate attacks with increasing intensity, i.e. up to $20\%$ of the nodes or edges in the network.  
When removing nodes, we consider the following cases: i) random deletion, ii) preferential removal of high-degree nodes, and iii) preferential removal of low-degree nodes. Preferential attacks are performed by selecting nodes with a probability $p$ proportional to their degree $k$, i.e. $p \propto k$ for high-degree nodes and $p \propto -k$ for low-degree nodes.
When perturbing edges, we test: i) random addition, ii) random deletion, and iii) random rewiring.
For each case, we simulated $1000$ repetitions and we computed the target control centrality $\tau$ for the driver nodes identified in the original network.
Then, we report the percentage of nodes that cease to be drivers  (i.e. $\tau=0$), that is, the percentage of nodes that are drivers in our analysis, but are no longer able to control any target in the perturbed case.

\section*{Acknowledgments}

We would like to thank Prof. Albert-Laszlo Barabasi for his helpful comments and suggestions. 
The research leading to these results has received funding from the French government under management of Agence Nationale de la Recherche as part of the "Investissements d'avenir" program, reference ANR-19-P3IA-0001 (PRAIRIE 3IA Institute) and reference ANR-10-IAIHU-06 (Agence Nationale de la Recherche-10-IA Institut Hospitalo-Universitaire-6), and from the Inria Project Lab Program (project Neuromarkers).
The content is solely the responsibility of the authors and does not necessarily represent the official views of any of the funding agencies.

\nolinenumbers

\clearpage
\bibliographystyle{plos2015.bst}
\nolinenumbers

\begin{small}
\bibliography{Exported-Items.bib}

\begin{thebibliography}{10}

\bibitem{jeong_lethality_2001}
Jeong H, Mason SP, Barab{\'a}si AL, Oltvai ZN.
\newblock Lethality and Centrality in Protein Networks.
\newblock Nature. 2001;411(6833):41--42.
\newblock doi:{10.1038/35075138}.

\bibitem{bonifazi_gabaergic_2009}
Bonifazi P, Goldin M, Picardo MA, Jorquera I, Cattani A, Bianconi G, et~al.
\newblock {{GABAergic Hub Neurons Orchestrate Synchrony}} in {{Developing
  Hippocampal Networks}}.
\newblock Science. 2009;326(5958):1419--1424.
\newblock doi:{10.1126/science.1175509}.

\bibitem{newman_networks:_2010}
Newman M.
\newblock Networks: {{An Introduction}}.
\newblock {Oxford, New York}: {Oxford University Press}; 2010.

\bibitem{rugh_linear_1995}
Rugh WJ, Kailath T.
\newblock Linear {{System Theory}}, 2nd {{Edition}}.
\newblock 2nd ed. {Upper Saddle River, NJ}: {Pearson}; 1995.

\bibitem{sontag_mathematical_1998}
Sontag ED.
\newblock Mathematical {{Control Theory}}: {{Deterministic Finite Dimensional
  Systems}}.
\newblock 2nd ed. Texts in {{Applied Mathematics}}. {New York}:
  {Springer-Verlag}; 1998.

\bibitem{wagner_number_2007}
Wagner SG.
\newblock On the Number of Matchings of a Tree.
\newblock European Journal of Combinatorics. 2007;28(4):1322--1330.
\newblock doi:{10.1016/j.ejc.2006.01.014}.

\bibitem{heuberger_number_2011}
Heuberger C, Wagner S.
\newblock The Number of Maximum Matchings in a Tree.
\newblock Discrete Mathematics. 2011;311(21):2512--2542.
\newblock doi:{10.1016/j.disc.2011.07.028}.

\bibitem{kalman_mathematical_1963}
Kalman RE.
\newblock Mathematical {{Description}} of {{Linear Dynamical Systems}}.
\newblock Journal of the Society for Industrial and Applied Mathematics Series
  A Control. 1963;1(2):152--192.
\newblock doi:{10.1137/0301010}.

\bibitem{hautus_stabilization_1970}
Hautus MLJ.
\newblock Stabilization Controllability and Observability of Linear Autonomous
  Systems.
\newblock Indagationes Mathematicae (Proceedings). 1970;73:448--455.
\newblock doi:{10.1016/S1385-7258(70)80049-X}.

\bibitem{lin_structural_1974}
Lin CT.
\newblock Structural Controllability.
\newblock IEEE Transactions on Automatic Control. 1974;19(3):201--208.
\newblock doi:{10.1109/TAC.1974.1100557}.

\bibitem{shields_structural_1976}
Shields R, Pearson J.
\newblock Structural Controllability of Multiinput Linear Systems.
\newblock IEEE Transactions on Automatic Control. 1976;21(2):203--212.
\newblock doi:{10.1109/TAC.1976.1101198}.

\bibitem{liu_controllability_2011}
Liu YY, Slotine JJ, Barab{\'a}si AL.
\newblock Controllability of Complex Networks.
\newblock Nature. 2011;473(7346):167--173.
\newblock doi:{10.1038/nature10011}.

\bibitem{hopcroft_n5/2_1971}
Hopcroft JE, Karp RM.
\newblock A N5/2 Algorithm for Maximum Matchings in Bipartite.
\newblock In: 12th {{Annual Symposium}} on {{Switching}} and {{Automata
  Theory}} (Swat 1971); 1971. p. 122--125.

\bibitem{uno_algorithms_1997}
Uno T.
\newblock Algorithms for Enumerating All Perfect, Maximum and Maximal Matchings
  in Bipartite Graphs.
\newblock In: Leong HW, Imai H, Jain S, editors. Algorithms and
  {{Computation}}. Lecture {{Notes}} in {{Computer Science}}. {Berlin,
  Heidelberg}: {Springer}; 1997. p. 92--101.

\bibitem{liu_number_2004}
Liu Y, Liu G.
\newblock Number of Maximum Matchings of Bipartite Graphs with Positive
  Surplus.
\newblock Discrete Mathematics. 2004;274(1):311--318.
\newblock doi:{10.1016/S0012-365X(03)00204-8}.

\bibitem{zdeborova_number_2006}
Zdeborov{\'a} L, M{\'e}zard M.
\newblock The Number of Matchings in Random Graphs.
\newblock Journal of Statistical Mechanics: Theory and Experiment.
  2006;2006(05):P05003--P05003.
\newblock doi:{10.1088/1742-5468/2006/05/P05003}.

\bibitem{klickstein_control_2019}
Klickstein IS, Sorrentino F.
\newblock Control {{Distance}} and {{Energy Scaling}} of {{Complex Networks}}.
\newblock IEEE Transactions on Network Science and Engineering. 2019; p. 1--1.
\newblock doi:{10.1109/TNSE.2018.2887042}.

\bibitem{gao_target_2014}
Gao J, Liu YY, D'Souza RM, Barab{\'a}si AL.
\newblock Target Control of Complex Networks.
\newblock Nature Communications. 2014;5:5415.
\newblock doi:{10.1038/ncomms6415}.

\bibitem{czeizler_target_2016}
Czeizler E, Gratie C, Chiu WK, Kanhaiya K, Petre I.
\newblock Target {{Controllability}} of {{Linear Networks}}.
\newblock In: Bartocci E, Lio P, Paoletti N, editors. Computational {{Methods}}
  in {{Systems Biology}}. Lecture {{Notes}} in {{Computer Science}}. {Cham}:
  {Springer International Publishing}; 2016. p. 67--81.

\bibitem{zhang_efficient_2017}
Zhang X, Wang H, Lv T.
\newblock Efficient Target Control of Complex Networks Based on Preferential
  Matching.
\newblock PLoS ONE. 2017;12(4).
\newblock doi:{10.1371/journal.pone.0175375}.

\bibitem{li_structural_2018-1}
Li J, Chen X, Pequito S, Pappas GJ, Preciado VM.
\newblock Structural {{Target Controllability}} of {{Undirected Networks}}.
\newblock In: 2018 {{IEEE Conference}} on {{Decision}} and {{Control}}
  ({{CDC}}). {Miami Beach, FL}: {IEEE}; 2018. p. 6656--6661.

\bibitem{lodish_gene_2000}
Lodish H, Berk A, Zipursky SL, Matsudaira P, Baltimore D, Darnell J.
\newblock Gene {{Replacement}} and {{Transgenic Animals}}.
\newblock Molecular Cell Biology 4th edition. 2000;.

\bibitem{hallett_transcranial_2000}
Hallett M.
\newblock Transcranial Magnetic Stimulation and the Human Brain.
\newblock Nature. 2000;406(6792):147--150.
\newblock doi:{10.1038/35018000}.

\bibitem{liu_control_2012}
Liu YY, Slotine JJ, Barab{\'a}si AL.
\newblock Control {{Centrality}} and {{Hierarchical Structure}} in {{Complex
  Networks}}.
\newblock PLOS ONE. 2012;7(9):e44459.
\newblock doi:{10.1371/journal.pone.0044459}.

\bibitem{gu_controllability_2015}
Gu S, Pasqualetti F, Cieslak M, Telesford QK, Yu AB, Kahn AE, et~al.
\newblock Controllability of Structural Brain Networks.
\newblock Nature Communications. 2015;6:8414.
\newblock doi:{10.1038/ncomms9414}.

\bibitem{murota_note_1990}
Murota K, Poljak S.
\newblock Note on a Graph-Theoretic Criterion for Structural Output
  Controllability.
\newblock IEEE Transactions on Automatic Control. 1990;35(8):939--942.
\newblock doi:{10.1109/9.58507}.

\bibitem{commault_functional_2019-1}
Commault C, {Van der Woude} J, Frasca P.
\newblock Functional Target Controllability of Networks: Structural Properties
  and Efficient Algorithms.
\newblock IEEE Transactions on Network Science and Engineering. 2019; p. 1--1.
\newblock doi:{10.1109/TNSE.2019.2937404}.

\bibitem{biggs_algebraic_1974}
Biggs N.
\newblock Algebraic {{Graph Theory}}.
\newblock {Cambridge University Press}; 1974.

\bibitem{robert_macrophages.com_2011}
Robert C, Lu X, Law A, Freeman TC, Hume DA.
\newblock Macrophages.Com: An on-Line Community Resource for Innate Immunity
  Research.
\newblock Immunobiology. 2011;216(11):1203--1211.
\newblock doi:{10.1016/j.imbio.2011.07.025}.

\bibitem{raza_logic-based_2008}
Raza S, Robertson KA, Lacaze PA, Page D, Enright AJ, Ghazal P, et~al.
\newblock A Logic-Based Diagram of Signalling Pathways Central to Macrophage
  Activation.
\newblock BMC Systems Biology. 2008;2:36.
\newblock doi:{10.1186/1752-0509-2-36}.

\bibitem{bitsch_tumour_2000}
Bitsch A, Kuhlmann T, Costa CD, Bunkowski S, Polak T, Br{\"u}ck W.
\newblock Tumour Necrosis Factor Alpha {{mRNA}} Expression in Early Multiple
  Sclerosis Lesions: {{Correlation}} with Demyelinating Activity and
  Oligodendrocyte Pathology.
\newblock Glia. 2000;29(4):366--375.
\newblock
  doi:{10.1002/(SICI)1098-1136(20000215)29:4<366::AID-GLIA7>3.0.CO;2-Y}.

\bibitem{mccormick_regulation_2015}
McCormick SM, Heller NM.
\newblock Regulation of {{Macrophage}}, {{Dendritic Cell}}, and {{Microglial
  Phenotype}} and {{Function}} by the {{SOCS Proteins}}.
\newblock Frontiers in Immunology. 2015;6.
\newblock doi:{10.3389/fimmu.2015.00549}.

\bibitem{koper_cxcl9_2018}
Koper OM, Kami{\'n}ska J, Sawicki K, Kemona H.
\newblock {{CXCL9}}, {{CXCL10}}, {{CXCL11}}, and Their Receptor ({{CXCR3}}) in
  Neuroinflammation and Neurodegeneration.
\newblock Advances in Clinical and Experimental Medicine: Official Organ
  Wroclaw Medical University. 2018;27(6):849--856.
\newblock doi:{10.17219/acem/68846}.

\bibitem{liu_interferon_2005}
Liu J, Guan X, Ma X.
\newblock Interferon Regulatory Factor 1 Is an Essential and Direct
  Transcriptional Activator for Interferon \{gamma\}-Induced
  {{RANTES}}/{{CCl5}} Expression in Macrophages.
\newblock The Journal of Biological Chemistry. 2005;280(26):24347--24355.
\newblock doi:{10.1074/jbc.M500973200}.

\bibitem{mosser_exploring_2008}
Mosser DM, Edwards JP.
\newblock Exploring the Full Spectrum of Macrophage Activation.
\newblock Nature Reviews Immunology. 2008;8(12):958--969.
\newblock doi:{10.1038/nri2448}.

\bibitem{barabasi_network_2004}
Barab{\'a}si AL, Oltvai ZN.
\newblock Network Biology: Understanding the Cell's Functional Organization.
\newblock Nature Reviews Genetics. 2004;5(2):101--113.
\newblock doi:{10.1038/nrg1272}.

\bibitem{pu_robustness_2012}
Pu CL, Pei WJ, Michaelson A.
\newblock Robustness Analysis of Network Controllability.
\newblock Physica A: Statistical Mechanics and its Applications.
  2012;391(18):4420--4425.
\newblock doi:{10.1016/j.physa.2012.04.019}.

\bibitem{wilson_socs_2014}
Wilson HM.
\newblock {{SOCS Proteins}} in {{Macrophage Polarization}} and {{Function}}.
\newblock Frontiers in Immunology. 2014;5.
\newblock doi:{10.3389/fimmu.2014.00357}.

\bibitem{liu_control_2016}
Liu YY, Barab{\'a}si AL.
\newblock Control {{Principles}} of {{Complex Networks}}.
\newblock Reviews of Modern Physics. 2016;88(3).
\newblock doi:{10.1103/RevModPhys.88.035006}.

\bibitem{lugagne_balancing_2017}
Lugagne JB, Sosa~Carrillo S, Kirch M, K{\"o}hler A, Batt G, Hersen P.
\newblock Balancing a Genetic Toggle Switch by Real-Time Feedback Control and
  Periodic Forcing.
\newblock Nature Communications. 2017;8.
\newblock doi:{10.1038/s41467-017-01498-0}.

\bibitem{sun_controllability_2016}
Sun X, Hu F, Wu S, Qiu X, Linel P, Wu H.
\newblock Controllability and Stability Analysis of Large Transcriptomic
  Dynamic Systems for Host Response to Influenza Infection in Human.
\newblock Infectious Disease Modelling. 2016;1(1):52--70.
\newblock doi:{10.1016/j.idm.2016.07.002}.

\bibitem{wuchty_controllability_2014}
Wuchty S.
\newblock Controllability in Protein Interaction Networks.
\newblock Proceedings of the National Academy of Sciences of the United States
  of America. 2014;111(19):7156--7160.
\newblock doi:{10.1073/pnas.1311231111}.

\bibitem{tang_developmental_2017}
Tang E, Giusti C, Baum GL, Gu S, Pollock E, Kahn AE, et~al.
\newblock Developmental Increases in White Matter Network Controllability
  Support a Growing Diversity of Brain Dynamics.
\newblock Nature Communications. 2017;8(1):1252.
\newblock doi:{10.1038/s41467-017-01254-4}.

\bibitem{dosenbach_distinct_2007}
Dosenbach NUF, Fair DA, Miezin FM, Cohen AL, Wenger KK, Dosenbach RAT, et~al.
\newblock Distinct Brain Networks for Adaptive and Stable Task Control in
  Humans.
\newblock Proceedings of the National Academy of Sciences.
  2007;104(26):11073--11078.
\newblock doi:{10.1073/pnas.0704320104}.

\bibitem{wagner_noninvasive_2007}
Wagner T, {Valero-Cabre} A, {Pascual-Leone} A.
\newblock Noninvasive {{Human Brain Stimulation}}.
\newblock Annual review of biomedical engineering. 2007;9:527--65.
\newblock doi:{10.1146/annurev.bioeng.9.061206.133100}.

\bibitem{menara_structural_2017}
Menara T, Gu S, Bassett DS, Pasqualetti F.
\newblock On {{Structural Controllability}} of {{Symmetric}} ({{Brain}})
  {{Networks}}.
\newblock arXiv:170605120 [cs, math]. 2017;.

\bibitem{gu_optimal_2017}
Gu S, Betzel RF, Mattar MG, Cieslak M, Delio PR, Grafton ST, et~al.
\newblock Optimal Trajectories of Brain State Transitions.
\newblock NeuroImage. 2017;148(Supplement C):305--317.
\newblock doi:{10.1016/j.neuroimage.2017.01.003}.

\bibitem{betzel_optimally_2016}
Betzel RF, Gu S, Medaglia JD, Pasqualetti F, Bassett DS.
\newblock Optimally Controlling the Human Connectome: The Role of Network
  Topology.
\newblock Scientific Reports. 2016;6.
\newblock doi:{10.1038/srep30770}.

\bibitem{muldoon_stimulation-based_2016}
Muldoon SF, Pasqualetti F, Gu S, Cieslak M, Grafton ST, Vettel JM, et~al.
\newblock Stimulation-{{Based Control}} of {{Dynamic Brain Networks}}.
\newblock PLOS Computational Biology. 2016;12(9):e1005076.
\newblock doi:{10.1371/journal.pcbi.1005076}.

\bibitem{klickstein_energy_2017}
Klickstein I, Shirin A, Sorrentino F.
\newblock Energy Scaling of Targeted Optimal Control of Complex Networks.
\newblock Nature Communications. 2017;8(1):15145.
\newblock doi:{10.1038/ncomms15145}.

\bibitem{zanudo_structure-based_2017}
Za{\~n}udo JGT, Yang G, Albert R.
\newblock Structure-Based Control of Complex Networks with Nonlinear Dynamics.
\newblock Proceedings of the National Academy of Sciences.
  2017;doi:{10.1073/pnas.1617387114}.

\bibitem{bullmore_complex_2009}
Bullmore E, Sporns O.
\newblock Complex Brain Networks: Graph Theoretical Analysis of Structural and
  Functional Systems.
\newblock Nature Reviews Neuroscience. 2009;10(3):186--198.
\newblock doi:{10.1038/nrn2575}.

\bibitem{bullmore_brain_2011}
Bullmore ET, Bassett DS.
\newblock Brain {{Graphs}}: {{Graphical Models}} of the {{Human Brain
  Connectome}}.
\newblock Annual Review of Clinical Psychology. 2011;7(1):113--140.
\newblock doi:{10.1146/annurev-clinpsy-040510-143934}.

\bibitem{de_vico_fallani_graph_2014}
De~Vico~Fallani F, Richiardi J, Chavez M, Achard S.
\newblock Graph Analysis of Functional Brain Networks: Practical Issues in
  Translational Neuroscience.
\newblock Philosophical Transactions of the Royal Society B: Biological
  Sciences. 2014;369(1653):20130521.
\newblock doi:{10.1098/rstb.2013.0521}.

\bibitem{taylor_dynamic_2009}
Taylor IW, Linding R, {Warde-Farley} D, Liu Y, Pesquita C, Faria D, et~al.
\newblock Dynamic Modularity in Protein Interaction Networks Predicts Breast
  Cancer Outcome.
\newblock Nature Biotechnology. 2009;27(2):199--204.
\newblock doi:{10.1038/nbt.1522}.

\bibitem{maslov_specificity_2002}
Maslov S, Sneppen K.
\newblock Specificity and {{Stability}} in {{Topology}} of {{Protein
  Networks}}.
\newblock Science. 2002;296(5569):910--913.
\newblock doi:{10.1126/science.1065103}.

\bibitem{drier_pathway-based_2013}
Drier Y, Sheffer M, Domany E.
\newblock Pathway-Based Personalized Analysis of Cancer.
\newblock Proceedings of the National Academy of Sciences.
  2013;110(16):6388--6393.
\newblock doi:{10.1073/pnas.1219651110}.

\bibitem{menche_disease_2015}
Menche J, Sharma A, Kitsak M, Ghiassian SD, Vidal M, Loscalzo J, et~al.
\newblock Disease Networks. {{Uncovering}} Disease-Disease Relationships
  through the Incomplete Interactome.
\newblock Science (New York, NY). 2015;347(6224):1257601.
\newblock doi:{10.1126/science.1257601}.

\bibitem{tong_novel_2017}
Tong T, Gao Q, Guerrero R, Ledig C, Chen L, Rueckert D.
\newblock A Novel Grading Biomarker for the Prediction of Conversion from Mild
  Cognitive Impairment to {{Alzheimer}}'s Disease.
\newblock IEEE Transactions on Biomedical Engineering. 2017;64(1):155--165.
\newblock doi:{10.1109/TBME.2016.2549363}.

\bibitem{uygun_utility_2016}
Uygun S, Peng C, {Lehti-Shiu} MD, Last RL, Shiu SH.
\newblock Utility and {{Limitations}} of {{Using Gene Expression Data}} to
  {{Identify Functional Associations}}.
\newblock PLoS Computational Biology. 2016;12(12).
\newblock doi:{10.1371/journal.pcbi.1005244}.

\bibitem{henry_bipom_2017}
Henry VJ, Sa{\"i}s F, Marchadier E, Dibie J, Goelzer A, Fromion V.
\newblock {{BiPOm}}: {{Biological}} Interlocked {{Process Ontology}} for
  Metabolism. {{How}} to Infer Molecule Knowledge from Biological Process?
\newblock In: International {{Conference}} on {{Biomedical Ontology}}, {{ICBO}}
  2017. {Newcastle upon Tyne, United Kingdom}; 2017. p.~np.

\bibitem{ruths_control_2014}
Ruths J, Ruths D.
\newblock Control {{Profiles}} of {{Complex Networks}}.
\newblock Science. 2014;343(6177):1373--1376.
\newblock doi:{10.1126/science.1242063}.

\bibitem{fallani_topological_2017}
Fallani FDV, Latora V, Chavez M.
\newblock A {{Topological Criterion}} for {{Filtering Information}} in
  {{Complex Brain Networks}}.
\newblock PLOS Computational Biology. 2017;13(1):e1005305.
\newblock doi:{10.1371/journal.pcbi.1005305}.

\bibitem{bentley_accurate_2008}
Bentley DR, Balasubramanian S, Swerdlow HP, Smith GP, Milton J, Brown CG,
  et~al.
\newblock Accurate Whole Human Genome Sequencing Using Reversible Terminator
  Chemistry.
\newblock Nature. 2008;456(7218):53--59.
\newblock doi:{10.1038/nature07517}.

\bibitem{mullighan_genome-wide_2007}
Mullighan CG, Goorha S, Radtke I, Miller CB, {Coustan-Smith} E, Dalton JD,
  et~al.
\newblock Genome-Wide Analysis of Genetic Alterations in Acute Lymphoblastic
  Leukaemia.
\newblock Nature. 2007;446(7137):758--764.
\newblock doi:{10.1038/nature05690}.

\bibitem{hauser_multiple_2015}
Hauser SL, Oksenberg JR, Baranzini SE.
\newblock Multiple {{Sclerosis}}.
\newblock In: Rosenberg's {{Molecular}} and {{Genetic Basis}} of
  {{Neurological}} and {{Psychiatric Disease}}. {Elsevier}; 2015. p.
  1001--1014.

\bibitem{chu_roles_2018}
Chu F, Shi M, Zheng C, Shen D, Zhu J, Zheng X, et~al.
\newblock The Roles of Macrophages and Microglia in Multiple Sclerosis and
  Experimental Autoimmune Encephalomyelitis.
\newblock Journal of Neuroimmunology. 2018;318:1--7.
\newblock doi:{10.1016/j.jneuroim.2018.02.015}.

\bibitem{strauss_immunophenotype_2015}
Strauss O, Dunbar PR, Bartlett A, Phillips A.
\newblock The Immunophenotype of Antigen Presenting Cells of the Mononuclear
  Phagocyte System in Normal Human Liver \textendash{} {{A}} Systematic Review.
\newblock Journal of Hepatology. 2015;62(2):458--468.
\newblock doi:{10.1016/j.jhep.2014.10.006}.

\bibitem{airas_hormonal_2015}
Airas L.
\newblock Hormonal and Gender-Related Immune Changes in Multiple Sclerosis.
\newblock Acta Neurologica Scandinavica. 2015;132(S199):62--70.
\newblock doi:{10.1111/ane.12433}.

\bibitem{song_comparison_2012}
Song L, Langfelder P, Horvath S.
\newblock Comparison of Co-Expression Measures: Mutual Information,
  Correlation, and Model Based Indices.
\newblock BMC Bioinformatics. 2012;13(1):328.
\newblock doi:{10.1186/1471-2105-13-328}.

\bibitem{steuer_mutual_2002}
Steuer R, Kurths J, Daub CO, Weise J, Selbig J.
\newblock The Mutual Information: {{Detecting}} and Evaluating Dependencies
  between Variables.
\newblock Bioinformatics. 2002;18(suppl\_2):S231--S240.

\bibitem{choi_differential_2005}
Choi JK, Yu U, Yoo OJ, Kim S.
\newblock Differential Coexpression Analysis Using Microarray Data and Its
  Application to Human Cancer.
\newblock Bioinformatics. 2005;21(24):4348--4355.
\newblock doi:{10.1093/bioinformatics/bti722}.

\bibitem{zhao_ranking_2011}
Zhao J, Yang TH, Huang Y, Holme P.
\newblock Ranking {{Candidate Disease Genes}} from {{Gene Expression}} and
  {{Protein Interaction}}: {{A Katz}}-{{Centrality Based Approach}}.
\newblock PLOS ONE. 2011;6(9):e24306.
\newblock doi:{10.1371/journal.pone.0024306}.

\bibitem{liseron-monfils_necorr_2018}
{Liseron-Monfils} C, Olson A, Ware D.
\newblock {{NECorr}}, a {{Tool}} to {{Rank Gene Importance}} in {{Biological
  Processes}} Using {{Molecular Networks}} and {{Transcriptome Data}}.
\newblock bioRxiv. 2018; p. 326868.
\newblock doi:{10.1101/326868}.

\bibitem{lohmann_eigenvector_2010}
Lohmann G, Margulies DS, Horstmann A, Pleger B, Lepsien J, Goldhahn D, et~al.
\newblock Eigenvector {{Centrality Mapping}} for {{Analyzing Connectivity
  Patterns}} in {{fMRI Data}} of the {{Human Brain}}.
\newblock PLOS ONE. 2010;5(4):e10232.
\newblock doi:{10.1371/journal.pone.0010232}.

\bibitem{sen_ranking_2019}
Sen B, Chu SH, Parhi KK.
\newblock Ranking {{Regions}}, {{Edges}} and {{Classifying Tasks}} in
  {{Functional Brain Graphs}} by {{Sub}}-{{Graph Entropy}}.
\newblock Scientific Reports. 2019;9(1):1--20.
\newblock doi:{10.1038/s41598-019-44103-8}.

\bibitem{li_fundamental_2016}
Li A, Cornelius SP, Liu YY, Wang L, Barab{\'a}si AL.
\newblock The Fundamental Advantages of Temporal Networks.
\newblock arXiv:160706168 [nlin]. 2016;.

\bibitem{zhang_controllability_2017}
Zhang Y, Garas A, Scholtes I.
\newblock Controllability of Temporal Networks: {{An}} Analysis Using
  Higher-Order Networks.
\newblock arXiv:170106331 [physics]. 2017;.

\bibitem{motik_structured_2008}
Motik B, Cuenca~Grau B, Sattler U.
\newblock Structured Objects in Owl: Representation and Reasoning.
\newblock In: Proceeding of the 17th International Conference on {{World Wide
  Web}} - {{WWW}} '08. {Beijing, China}: {ACM Press}; 2008. p. 555.

\bibitem{musen_protege_2015}
Musen MA.
\newblock The {{Prot\'eg\'e Project}}: {{A Look Back}} and a {{Look Forward}}.
\newblock AI matters. 2015;1(4):4--12.
\newblock doi:{10.1145/2757001.2757003}.

\bibitem{agarwala_database_2018}
Agarwala R, Barrett T, Beck J, Benson DA, Bollin C, Bolton E, et~al.
\newblock Database Resources of the {{National Center}} for {{Biotechnology
  Information}}.
\newblock Nucleic Acids Research. 2018;46(D1):D8--D13.
\newblock doi:{10.1093/nar/gkx1095}.

\bibitem{the_uniprot_consortium_uniprot_2019}
Consortium TU.
\newblock {{UniProt}}: A Worldwide Hub of Protein Knowledge.
\newblock Nucleic Acids Research. 2019;47(D1):D506--D515.
\newblock doi:{10.1093/nar/gky1049}.

\bibitem{stelzer_genecards_2016}
Stelzer G, Rosen N, Plaschkes I, Zimmerman S, Twik M, Fishilevich S, et~al.
\newblock The {{GeneCards Suite}}: {{From Gene Data Mining}} to {{Disease
  Genome Sequence Analyses}}.
\newblock Current Protocols in Bioinformatics. 2016;54(1):1.30.1--1.30.33.
\newblock doi:{10.1002/cpbi.5}.

\bibitem{thompson_diagnosis_2018}
Thompson AJ, Banwell BL, Barkhof F, Carroll WM, Coetzee T, Comi G, et~al.
\newblock Diagnosis of Multiple Sclerosis: 2017 Revisions of the {{McDonald}}
  Criteria.
\newblock The Lancet Neurology. 2018;17(2):162--173.
\newblock doi:{10.1016/S1474-4422(17)30470-2}.

\bibitem{andrews_fastqc_2010}
Andrews S. {{FastQC}}: {{A Quality Control Tool}} for {{High Throughput
  Sequence Data}}; 2010.

\bibitem{li_sequence_2009}
Li H, Handsaker B, Wysoker A, Fennell T, Ruan J, Homer N, et~al.
\newblock The {{Sequence Alignment}}/{{Map}} Format and {{SAMtools}}.
\newblock Bioinformatics. 2009;25(16):2078--2079.
\newblock doi:{10.1093/bioinformatics/btp352}.

\bibitem{wang_rseqc_2012}
Wang L, Wang S, Li W.
\newblock {{RSeQC}}: Quality Control of {{RNA}}-Seq Experiments.
\newblock Bioinformatics. 2012;28(16):2184--2185.
\newblock doi:{10.1093/bioinformatics/bts356}.

\bibitem{dobin_star_2013}
Dobin A, Davis CA, Schlesinger F, Drenkow J, Zaleski C, Jha S, et~al.
\newblock {{STAR}}: Ultrafast Universal {{RNA}}-Seq Aligner.
\newblock Bioinformatics. 2013;29(1):15--21.
\newblock doi:{10.1093/bioinformatics/bts635}.

\bibitem{noli_discordant_2015}
Noli L, Capalbo A, Ogilvie C, Khalaf Y, Ilic D.
\newblock Discordant {{Growth}} of {{Monozygotic Twins Starts}} at the
  {{Blastocyst Stage}}: {{A Case Study}}.
\newblock Stem Cell Reports. 2015;5(6):946--953.
\newblock doi:{10.1016/j.stemcr.2015.10.006}.

\bibitem{love_moderated_2014}
Love MI, Huber W, Anders S.
\newblock Moderated Estimation of Fold Change and Dispersion for {{RNA}}-Seq
  Data with {{DESeq2}}.
\newblock Genome Biology. 2014;15(12).
\newblock doi:{10.1186/s13059-014-0550-8}.

\bibitem{r_core_team_r_2014}
Team RC. R: {{A Language}} and {{Environment}} for {{Statistical Computing}};
  2014.

\end{thebibliography}
\end{small}

\par\null
\section*{Figures}
\label{sec: figures}


\newpage

\newpage

\newpage

\newpage

\newpage
\renewcommand{\thefigure}{S\arabic{figure}}

\setcounter{figure}{0}

\newpage
\section*{Supplementary figures}
\label{sec: SM-fig}

\begin{figure}[H]
	\centering
	\includegraphics[width=\textwidth]{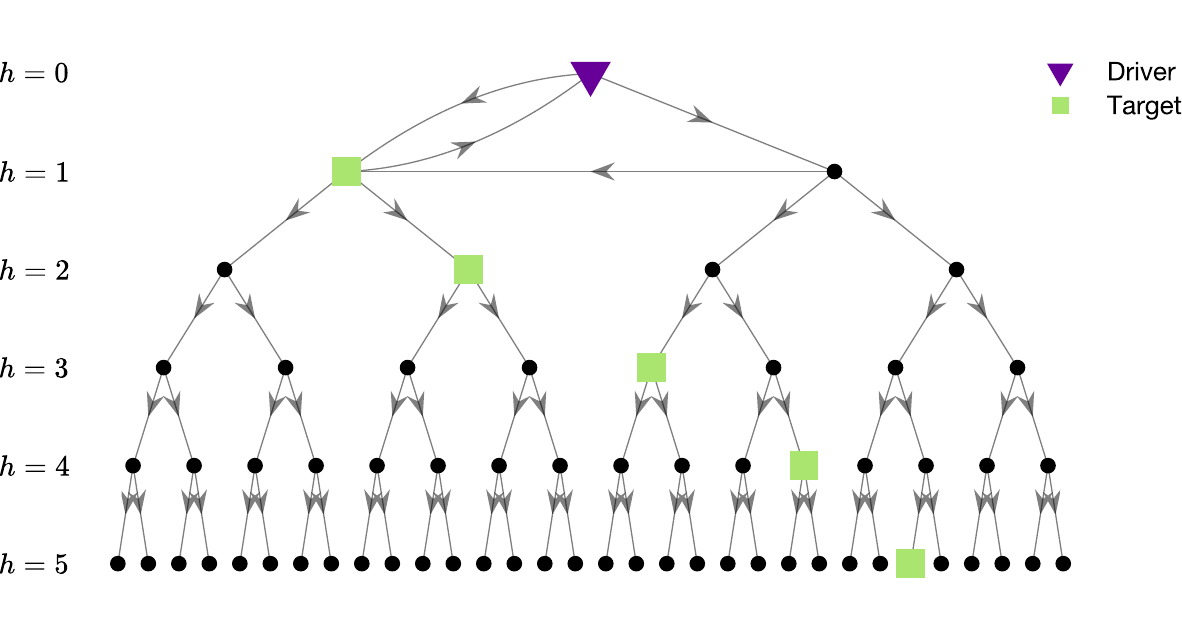}
	
	\caption{\small \textbf{Methodological validation of step-wise target controllability.} We start from a simple directed full binary tree, and we add a cycle among the first three nodes. The driver is the root of the tree, while we put a target node in each level of the tree. Target nodes are then ranked according to their height $h$. 
	This configuration is fully controllable by construction regardless of the tree's height $h$.
	However, for $h=5$ (i.e. $N=63$), the standard procedure computing the rank of the full Kalman controllability matrix cannot retrieve all the targets, as the rank computation is deficient due to numerical errors.
	Instead, by using the step-wise target controllability we can correctly identify the controllable targets up to $h=10$, i.e. $N=2047$. 
	}
\label{fig: binary tree}
\end{figure}

\newpage
\begin{figure}[H]
	\centering
	\includegraphics[width=\textwidth]{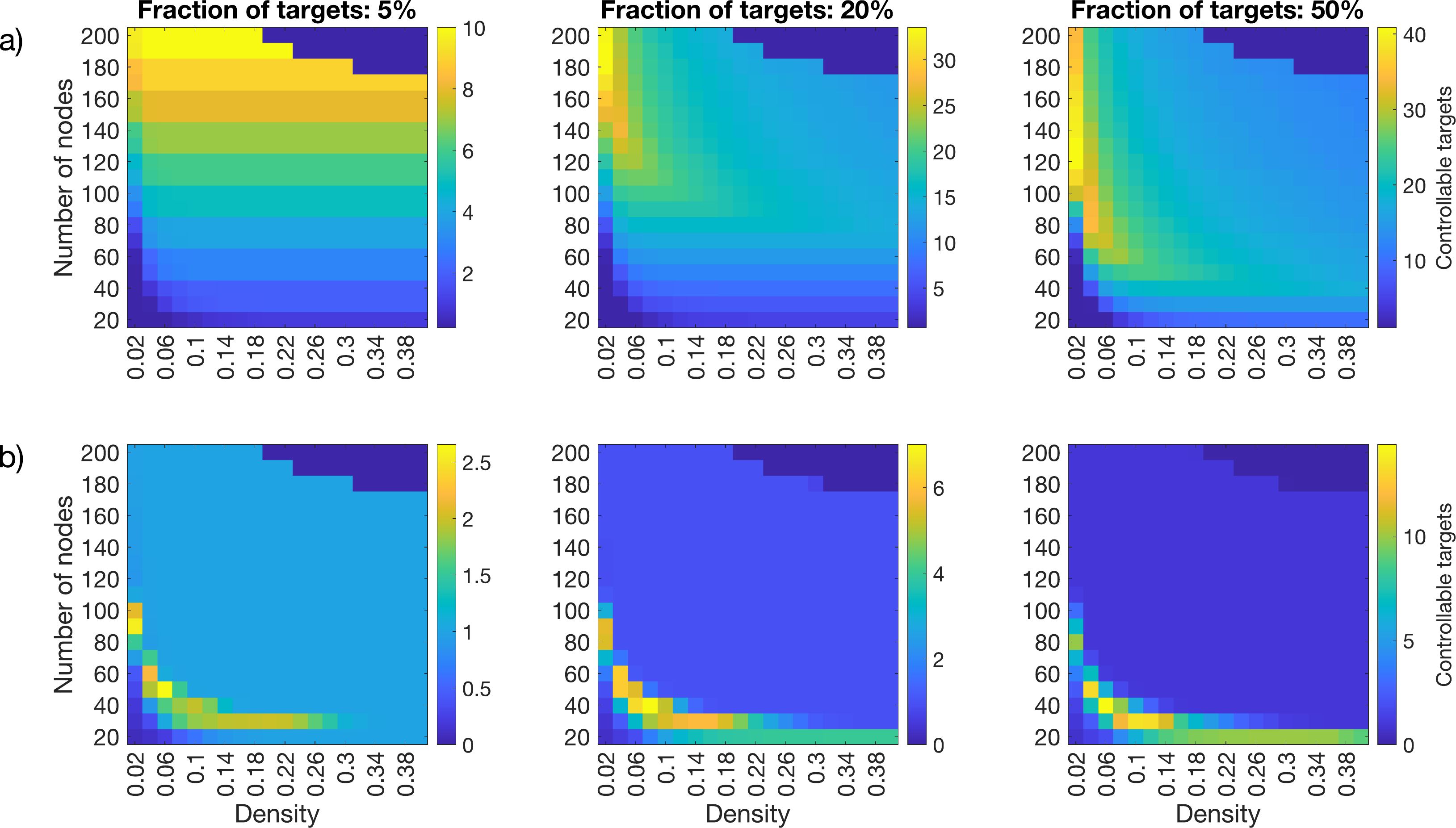}
	
	\caption{\small \textbf{Average number of targets that can be controlled by a single driver node}, computed on samples of 100 connected and directed random networks, at the varying of size and density of the networks, and target set size.
	 Panel \textbf{a)} corresponds to the step-wise target controllability.
	 Panel \textbf{b)} corresponds to a standard procedure that computes the rank of the full Kalman controllability matrix.
Results show that in general our method is able to retrieve a larger number of controllable targets as compared to the standard procedure. More specifically, when the target set contains $5\%$ of the network nodes, results are quite stable across different connection densities. For larger target-set sizes our method works better when the connection density is relatively low ($0.02$-$0.10$).
It is important also to notice that the computation of the rank starts to fail in correspondence of larger and denser networks (i.e., $N>180$ and density $> 0.20$).
	}
\label{fig: computational-complexity}
\end{figure}

\begin{figure}[H]
	\centering
	\includegraphics[width=\textwidth]{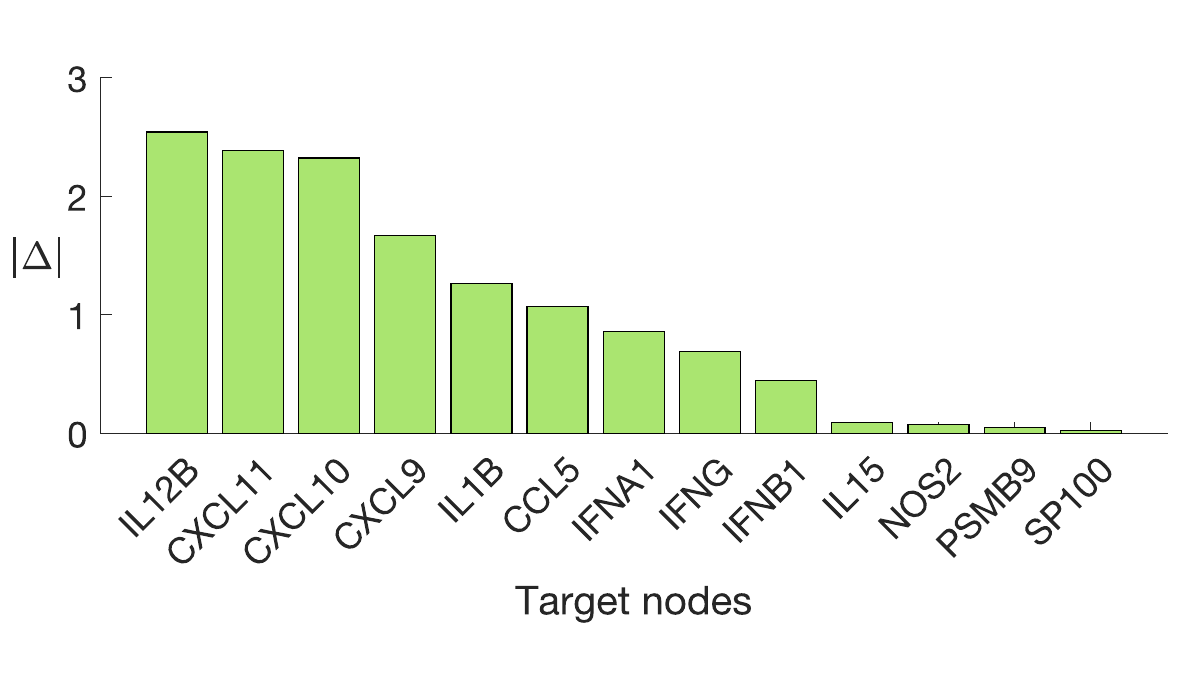}
	\caption{\small\textbf{Hierarchy among target genes.} Genes corresponding to the $13$ secreted molecules are ranked according to the absolute value of fold change $\Delta$ in the gene activation between the multiple sclerosis (MS) group and the healthy control (HC) group (\textbf{Materials and methods}).
}
\label{fig: ordered targets}
\end{figure}

	

\newpage
\begin{figure}[H]
	\centering
	\includegraphics[width=\textwidth]{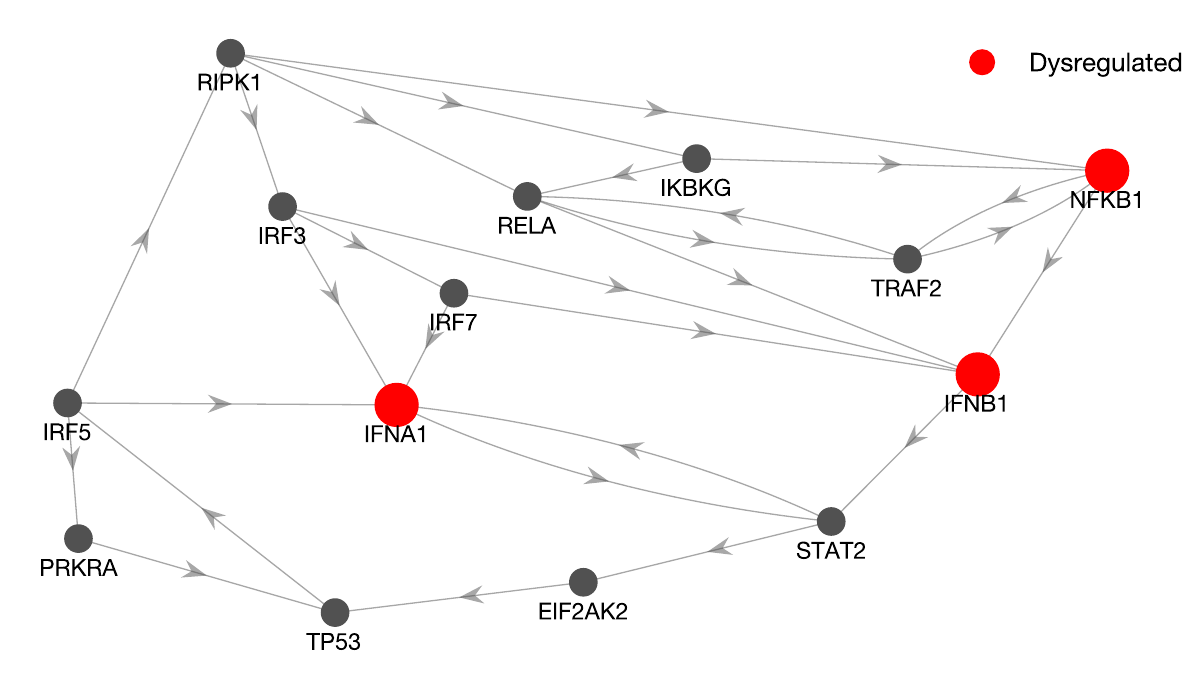}
	
	\caption{\small \textbf{Subnetwork illustrating the feedback cycle between dysregulated genes IFNA1, IFNB1 and NFKB1.} 
	The three nodes belong to the only strongly connected component (a subnetwork in which every node is reachable from any other node) of the network having more than two nodes.
	It plays, thus, a central role in the network topology.
	}
\label{fig: connected-component}
\end{figure}

\newpage
\section*{Supplementary tables}
\label{sec: SM-tab}

%
%
%
%

\begin{table}[H]
\caption{\small List of all node genes and associated class depending on their functional role.}
\label{table:s1-gene-class}
\end{table}

\begin{table}[H]
\caption{\small Distribution of driver nodes across different classes of genes. }
\label{table:s2-drivers-targets}
\end{table}

\begin{table}[H]
\caption{\small Spearman cross-correlation and p-values values for controllable driver-target node pairs. Values for pairs that were not controllable were replaced by not-a-number (NaN). Both values for multiple sclerosis patients (MS) and healthy controls (HC) are reported.}
\label{table:s3-correlation}
\end{table}

\begin{table}[H]
\caption{\small Log-transformed gene expression for multiple sclerosis patients (MS) and healthy controls (HC), in the `alert' (M0) and `pro-infalammatory' (M1) macrophage activation states; computation of the $\Delta$ as fold change.}
\label{table:s4-delta}
\end{table}

\end{document}